\documentclass[a4paper,twoside]{IEEEtran}

\usepackage[IEEEtran,notheorems,hyperrefurl]{research18}
\usepackage{xcolor}
\usepackage{import}

\newtheorem{theorem}{Theorem}
\newtheorem{lemma}[theorem]{Lemma}
\newtheorem{proposition}[theorem]{Proposition}
\newtheorem{corollary}[theorem]{Corollary}

\newtheorem{definition}[theorem]{Definition}
\newtheorem{remark}[theorem]{Remark}
\newtheorem{example}{Example}

\DeclareSymbolFont{symbolsC}{U}{txsyc}{m}{n}
\SetSymbolFont{symbolsC}{bold}{U}{txsyc}{bx}{n}
\DeclareFontSubstitution{U}{txsyc}{m}{n}
\DeclareMathSymbol{\medcirc}{\mathbin}{symbolsC}{7}
\newcommand{\sphere}{\medcirc}
\newcommand{\simplex}{\triangle}
\newcommand{\cube}{\square}
\newcommand{\octahedron}{\Diamond}
\newcommand{\vol}{\operatorname{vol}}
\newcommand{\Extr}{\operatorname{Extr}}
\renewcommand{\co}[1]{\operatorname{conv}\left(#1\right)} 

\renewcommand{\bigco}[1]{\operatorname{conv}\bigl(#1\bigr)}

\begin{document}
\sloppy

\title{Rate-Distortion Problems of the Poisson Process based on a
  Group-Theoretic Approach} 

\author{Hui-An Shen,
  Stefan M.~Moser,~\IEEEmembership{Senior Member,~IEEE}, and
  Jean-Pascal Pfister
  \thanks{This work was supported by the Swiss National Science
    Foundation grants PP00P3\_179060 (J.-P.~P.) and 31003A\_175644
    (H.-A.~S.).  It was presented in part at the 2021 IEEE Information
    Theory Workshop. \emph{(Corresponding author: Hui-An Shen.)}

    Hui-An Shen and Jean-Pascal Pfister belong to the Theoretical
    Neuroscience Group, University of Bern, Switzerland, and are
    affiliated with the Institute of Neuroinformatics, University of
    Zurich and ETH Zurich, Switzerland (e-mail: \{jeanpascal.pfister,
    huian.shen\}@unibe.ch).

    Stefan M. Moser is with the Signal and Information Processing Lab,
    ETH Zurich, Switzerland, and is affiliated with the Institute of
    Communications Engineering, National Yang Ming Chiao Tung
    University, Hsinchu, Taiwan (e-mail: moser@isi.ee.ethz.ch).}}
  
\maketitle

\begin{abstract}
  We study rate-distortion problems of a Poisson process using a group
  theoretic approach. By describing a realization of a Poisson point
  process with either point timings or inter-event (inter-point)
  intervals and by choosing appropriate distortion measures, we
  establish rate-distortion problems of a homogeneous Poisson process
  as ball- or sphere-covering problems for realizations of the
  hyperoctahedral group in $\Reals^n$. Specifically, the realizations
  we investigate are a hypercube and a hyperoctahedron. Thereby we
  unify three known rate-distortion problems of a Poisson process
  (with different distortion measures, but resulting in the same
  rate-distortion function) with the Laplacian-$\ell_1$
  rate-distortion problem.
\end{abstract}


\section{Introduction}
\label{sec:introduction}

In \cite{shenmoserpfister21_1} we have studied the well-known
geometric concept of \emph{sphere covering}, which beautifully
explains the rate-distortion problem for the Gaussian source under an
$\ell_2$-distortion measure, and we have applied it to a Poisson
process with an appropriately chosen distortion measure. We thereby
succeeded in providing a new geometric proof of the converse to a
rate-distortion theorem for a Poisson process.

In this work we would like to build on these insights and investigate
more in details what underlying geometric structure is required for
such a geometric proof. We will provide geometric, sphere-covering
proofs to four known different rate-distortion problems: three
rate-distortion problems for the Poisson process with different
distortion measures, and one rate-distortion problem for the
Laplacian-$\ell_1$ source. Thereby we rely on group theory and some of
its basic tools to connect these four different
problems. Specifically, we use groups to describe certain symmetries
inherent in the Poisson process.

The \emph{permutation group}---the group that contains all possible
permutations of $n$ objects---is ideal to describe possible
relabelings of event times of a realization of a homogeneous Poisson
process (conditional on a given number of events), which is possible
because the Poisson process is memoryless. So, we use the event
timings to describe the Poisson process, rely on the permutation group
to set it into a geometric viewpoint, and---using point-covering
distortion as a distortion measure---are then able to derive the
point-covering rate-distortion function via the sphere-covering
argument.

If, instead of the permutation group, its subgroup with only the
identity permutation is used, the geometric picture fits to the case
of the queueing rate-distortion problem. Note that both these problems
correspond to $\ell_{\infty}$-ball covering. They are explored in
Section~\ref{sec:rate-dist-ball}.

Similarly, the \emph{reflection group}---the group describing all
possible reflections on the $(n-1)$-dimensional principle hyperplanes
($x_i=0$) in the $n$-dimensional space---can be used to describe
symmetries in the context of the inter-event intervals. In this case,
we use the exponentially distributed inter-event intervals to describe
the homogeneous Poisson process, rely on the subgroup of the
reflection group containing only the identity element to obtain a
geometric picture for it, and---in combination with a corresponding
onesided $\ell_1$-distortion measure---are able to give the
sphere-covering picture for the exponential onesided
$\ell_1$-rate-distortion problem.

If we use the complete reflection group  in combination with the
$\ell_1$-distortion measure, we can geometrically represent the case
of the Laplacian $\ell_1$-rate-distortion problem. So, these two
problems are $\ell_1$-sphere-covering problems. They are explored in
Section~\ref{sec:rate-dist-sphere}. 

\begin{figure*}[htb]
    \centering
    \begin{tikzpicture}
        [scale=0.75,>={Stealth[scale=0.7]},
        box/.style={rectangle,draw,thick,inner sep=0pt,minimum height=1.5cm,
                minimum width=2cm,align=center}]

        \draw[ultra thick] (-12,0.2)--(12,0.2);
        \draw[ultra thick] (-12,5.8)--(12,5.8);
        \draw[ultra thick] (0,-5.3)--(0,7.2);

        \begin{scope}
            \draw[thick] (-8,6.5) to
            node[below,at start] {$0$}
            node[below,at end] {$T$}
            (-4,6.5);
            \draw[thick] (-8,6.4)--(-8,6.6);
            \draw[thick] (-4,6.4)--(-4,6.6);
            \draw[thick] (-7,6.5) to
            node[at start,below] {$t_1$}
            (-7,7);
            \draw[thick] (-6.5,6.5) to (-6.5,7);
            \node at (-6,6.15) {$\cdots$};
            \draw[thick] (-5,6.5) to
            node[at start,below] {$t_n$}
            (-5,7);
        \end{scope}

        \begin{scope}[xshift=12cm,yshift=-2mm]
            \draw[thick,->] (-8,7) -- (-8,6.5) -- (-4,6.5);
            \draw[thick] (-7,6.5) -- (-7,7);
            \draw[thick,<->] (-8,6.75) to
            node[midway,above] {$\tau_1$}
            (-7,6.75);
            \node at (-6.5,6.75) {$\cdots$};
            \draw[thick] (-6.1,6.5) -- (-6.1,7);
            \draw[thick,<->] (-6.1,6.75) to
            node[midway,above] {$\tau_n$}
            (-4.6,6.75);
            \draw[thick] (-4.6,6.5) -- (-4.6,7);
        \end{scope}

        \begin{scope}[yshift=-6.1cm]
            \node[right] at (-11.8,5.5) {subgroup $\{\sigma_1\}$ of $\const{S}_n$:};
            \node[right] at (-11.8,4.6) {$\left\{\begin{pmatrix} 1 & \cdots &
                n            \\ 1 & \cdots & n\end{pmatrix}\right\}$};
            \draw[ultra thick, red] (-11.9,3.7)--(-11.9,6)
            --(-7.2,6)--(-7.2,3.7)--(-11.9,3.7)--(-11.9,6);
        \end{scope}
        \begin{scope}[yshift=-5.6cm,xshift=2.8mm]
            \node[right] at (-6.5,5) {$n$-simplex};
            \node[right] at (-6.5,4.5) {\ssmall $\set{S}_{\sigma_1} = \bigl\{
                    \vect{t}\in\Reals^n\colon$};
            \node[right] at (-6.5,4) {\ssmall $\quad 0<t_{\sigma_1(1)}<\cdots$};
            \node[right] at (-6.5,3.5) {\ssmall $\quad <t_{\sigma_1(n)}<1 \bigr\}$};
            \begin{scope}[xshift=4mm]
                \draw[fill,blue!30] (-3.8,3.5)--(-3.8,5)--(-2.3,5);
                \draw (-3.8,3.5)--(-2.3,5);
                \draw[thick,->] (-4,3.5) to
                node[at end,right] {\scriptsize $t_1$}
                (-1.5,3.5);
                \draw[thick,->] (-3.8,3.3) to
                node[at end,left] {\scriptsize $t_2$}
                (-3.8,5.3);
                \draw[thick] (-3.8,5)--(-2.3,5)--(-2.3,3.5);
                \node at (-3.4,4.6) {$\set{S}_{\sigma_1}$};
                \draw[very thick, green!70!black] (-3.8,3.5)--(-3.8,3.9)
                --(-3.4,3.9)--(-3.8,3.5)--(-3.8,3.9);
                \node[green!70!black] at (-3.25,3.75) {\small $d_{\text{q}}$};
            \end{scope}
            \draw[ultra thick, blue] (-6.6,3.2)--(-6.6,5.5)
            --(-0.5,5.5)--(-0.5,3.2)--(-6.6,3.2)--(-6.6,5.5);
        \end{scope}
        \begin{scope}[yshift=-6.1cm]
            \node[right] at (-11.8,3) {RD: source: $\vect{t}'\sim\Uniform{[0,T]^n}$,
                \quad $\vect{t}=\text{sort}(\vect{t}')$};
            \node[right] at (-11.8,2) {\phantom{RD:}
                dist.: \scriptsize $d_{\textnormal{q}}(\vect{t},\hvect{x})
                    = \begin{cases}
                        \sum_{i=1}^{N_{\vect{t}}(T)} t_i -\max\{t_{i-1},\hat{x}_i\}
                               & \text{if } (**)
                        \\
                        \infty & \text{otherwise}\end{cases}$};
            \draw[ultra thick, green!70!black] (-11.9,1)--(-11.9,3.5)
            --(-0.2,3.5)--(-0.2,1)--(-11.9,1)--(-11.9,3.5);
        \end{scope}

        \begin{scope}[yshift=-5mm]
            \node[right] at (-11.8,5.5) {\textbf{permutation group}};
            \node[right] at (-11.8,4.9) {of all permutations:};
            \node[right] at (-11.8,4.3) {$\const{S}_n = \{\sigma_1, \ldots,
                    \sigma_{n!}\}$};
            \draw[ultra thick, red] (-11.9,3.7)--(-11.9,6)
            --(-7.2,6)--(-7.2,3.7)--(-11.9,3.7)--(-11.9,6);
            \node at (-6.75,4.9) {$\cong$};
        \end{scope}
        \begin{scope}[xshift=2.8mm]
            \node[right] at (-6.5,5) {$n$-cube};
            \node[right] at (-6.5,4.4) {$\cube^n$};
            \node[right] at (-6.5,3.8) {$\{\set{S}_{\sigma}\colon
                    \sigma\in\const{S}_n\}$};
            \begin{scope}[xshift=4mm]
                \draw[fill,blue!30] (-3.8,5)--(-2.3,5)--(-2.3,3.5)--(-3.8,3.5);
                \draw[thick,->] (-4,3.5) to
                node[at end,right] {\scriptsize $t_1$}
                (-1.5,3.5);
                \draw[thick,->] (-3.8,3.3) to
                node[at end,left] {\scriptsize $t_2$}
                (-3.8,5.3);
                \draw[thick] (-3.8,5)--(-2.3,5)--(-2.3,3.5);
                \draw[thick,densely dashed] (-3.8,3.5)--(-2.3,5);
                \node at (-3.4,4.6) {$\set{S}_{\sigma_1}$};
                \node at (-2.6,3.9) {$\set{S}_{\sigma_2}$};
                \draw[very thick, green!70!black] (-3.6,3.7) rectangle (-3.2,4.1);
                \node[green!70!black] at (-3.3,3.45) {\small $d_{\text{pc}}$};
            \end{scope}
            \draw[ultra thick, blue] (-6.6,3.2)--(-6.6,5.5)
            --(-0.5,5.5)--(-0.5,3.2)--(-6.6,3.2)--(-6.6,5.5);
        \end{scope}
        \begin{scope}[yshift=-5mm]
            \node[right] at (-11.8,3) {RD: source: $\vect{t}\sim\Uniform{[0,T]^n}$};
            \node[right] at (-11.8,2) {\phantom{RD:}
                dist.: \scriptsize $d_{\textnormal{pc}}(\vect{t},\hvect{x})
                    = \begin{cases}\D \int_0^T
                        \hat{x}(s)\dd s & \text{if } \D\int_0^T \hat{x}(s)
                        \sum_{i=1}^n \delta(t_i-s)\dd s = n                \\
                        \infty          & \text{otherwise}\end{cases}$};
            \draw[ultra thick, green!70!black] (-11.9,1)--(-11.9,3.5)
            --(-0.2,3.5)--(-0.2,1)--(-11.9,1)--(-11.9,3.5);
        \end{scope}

        \begin{scope}[xshift=12.1cm,yshift=-6.1cm]
            \node[right] at (-11.8,5.5) {subgroup $\{h_1\}$ of $\const{H}_n$:};
            \node[right] at (-11.8,4.5) {$\{( 1, 1, \ldots , 1)\}$};
            \draw[ultra thick, red] (-11.9,3.7)--(-11.9,6)
            --(-7.2,6)--(-7.2,3.7)--(-11.9,3.7)--(-11.9,6);
        \end{scope}
        \begin{scope}[xshift=12.38cm,yshift=-5.6cm]
            \node[right] at (-6.5,5) {$(n-1)$-};
            \node[right] at (-6.5,4.5) {simplex};
            \node[right] at (-6.5,3.7)
            {$\simplex^{n-1}\left(\frac{n}{\lambda}\right)$};
            \begin{scope}[xshift=4mm]
                \draw[thick,->] (-4,3.5) to
                node[pos=0.95,right] {\scriptsize $\tau_1$}
                (-1.5,3.5);
                \draw[thick,->] (-3.8,3.3) to
                node[pos=0.95,left] {\scriptsize $\tau_2$}
                (-3.8,5.3);
                \draw[very thick,blue] (-3.8,5)--(-2.3,3.5);
                \draw[ultra thick, green!70!black] (-3.41,4.6) -- (-3.12,4.31);
                \draw[ultra thick, green!70!black] (-3.405,4.605) -- (-3.115,4.315);
                \draw[thick, green!70!black] (-3.4,4.6) -- (-3.4,4.32)
                -- (-3.12,4.32); 
                \node[green!70!black] at (-2.9,4.55) {\small $d_{1}$};
            \end{scope}
            \draw[ultra thick, blue] (-6.6,3.2)--(-6.6,5.5)
            --(-0.5,5.5)--(-0.5,3.2)--(-6.6,3.2)--(-6.6,5.5);
        \end{scope}
        \begin{scope}[xshift=12.1cm,yshift=-6.1cm]
            \node[right] at (-11.8,3) {RD: source: $\tau_i' \sim \frac{1}{2}\lambda
                    \ope^{-\lambda|\tau|}, \quad \tau_i=\text{abs}(\tau_i')$  };
            \node[right] at (-11.8,2) {\phantom{RD:}
                dist.: \scriptsize $d_{1}(\vectg{\tau},\hvect{x})
                    = \begin{cases} \frac{\lambda}{n}
                        \sum_{i=1}^n \bigabs{\tau_i-\hat{x}_i}
                               & \text{if } \tau_i-\hat{x}_i \ge 0  \,\forall i \in [n] \\
                        \infty & \text{otherwise}\end{cases}$};
            \draw[ultra thick, green!70!black] (-11.9,1)--(-11.9,3.5)
            --(-0.2,3.5)--(-0.2,1)--(-11.9,1)--(-11.9,3.5);
        \end{scope}

        \begin{scope}[xshift=12.1cm,yshift=-5mm]
            \node[right] at (-11.8,5.5) {\textbf{reflection group}};
            \node[right] at (-11.8,4.95) {of all reflections:};
            \node[right] at (-11.8,4.3) {$\const{H}_n = \{-1,+1\}^n$};
            \draw[ultra thick, red] (-11.9,3.7)--(-11.9,6)
            --(-7.2,6)--(-7.2,3.7)--(-11.9,3.7)--(-11.9,6);
            \node at (-6.75,4.9) {$\cong$};
        \end{scope}
        \begin{scope}[xshift=12.38cm]
            \node[right] at (-6.5,5) {$\ell_1$-sphere};
            \node[right] at (-6.5,4.4)
            {$\sphere_1^{n-1}\left(\frac{n}{\lambda}\right)$};
            \begin{scope}[xshift=4mm]
                \draw[thick,->] (-4,4.3) to
                node[pos=0.95,below] {\scriptsize $\tau_1$}
                (-1.4,4.3);
                \draw[thick,->] (-2.7,3.3) to
                node[pos=0.95,left] {\scriptsize $\tau_2$}
                (-2.7,5.4);
                \draw[very thick,blue] (-2.7,5)--(-2,4.3)
                --(-2.7,3.6)--(-3.4,4.3)--(-2.7,5)--(-2,4.3);
                \draw[ultra thick, green!70!black] (-2.5,4.8) -- (-2.29,4.59);
                \draw[thick, green!70!black] (-2.61,4.69) -- (-2.39,4.91)
                -- (-2.18,4.69)--(-2.39,4.48)--(-2.61,4.69) -- (-2.39,4.91);
                \node[green!70!black] at (-1.8,5) {\small $d_{\text{norm}}$};
            \end{scope}
            \draw[ultra thick, blue] (-6.6,3.2)--(-6.6,5.5)
            --(-0.5,5.5)--(-0.5,3.2)--(-6.6,3.2)--(-6.6,5.5);
        \end{scope}
        \begin{scope}[xshift=12.1cm,yshift=-5mm]
            \node[right] at (-11.8,3) {RD: source:
                $\tau_i \sim \frac{1}{2}\lambda
                    \ope^{-\lambda|\tau|}$ \qquad\quad ($\text{sgn}(\tau_i)=\pm 1$)};
            \node[right] at (-11.8,2) {\phantom{RD:}
                dist.: \scriptsize $\D d_{\textnormal{norm}}(\vectg{\tau},\hvect{x})
                    = \frac{\lambda}{n} \sum_{i=1}^n \bigabs{\tau_i-\hat{x}_i}$};
            \draw[ultra thick, green!70!black] (-11.9,1)--(-11.9,3.5)
            --(-0.2,3.5)--(-0.2,1)--(-11.9,1)--(-11.9,3.5);
        \end{scope}


    \end{tikzpicture}
    \caption{Unification of four rate distortion problems. The left
      and right columns present the symmetries in the timing
      description (permutation group) and the interval description
      (reflection group), respectively. The upper and lower rows
      present the respective group and its subgroup. Each quadrant
      also illustrates its own rate-distortion problem with its source
      set (in the blue box) and its natural distortion measure ``dist''
      (in the green box). In the left lower quadrant, we use Cauchy's
      two-line notation for permutation to denote $\sigma_1$; and the
      (**) condition for finite distortion is
      $N_{\hvect{x}}(T) = N_{\vect{t}}(T)$ and
      $N_{\hvect{x}}(s) \ge N_{\vect{t}}(s) \,\forall s\in[0,T]$. Note
      that the blue lines on the right column illustrate the region
      where the source is concentrated (when $n\to\infty$).}
    \label{fig:bigpicture}
\end{figure*}
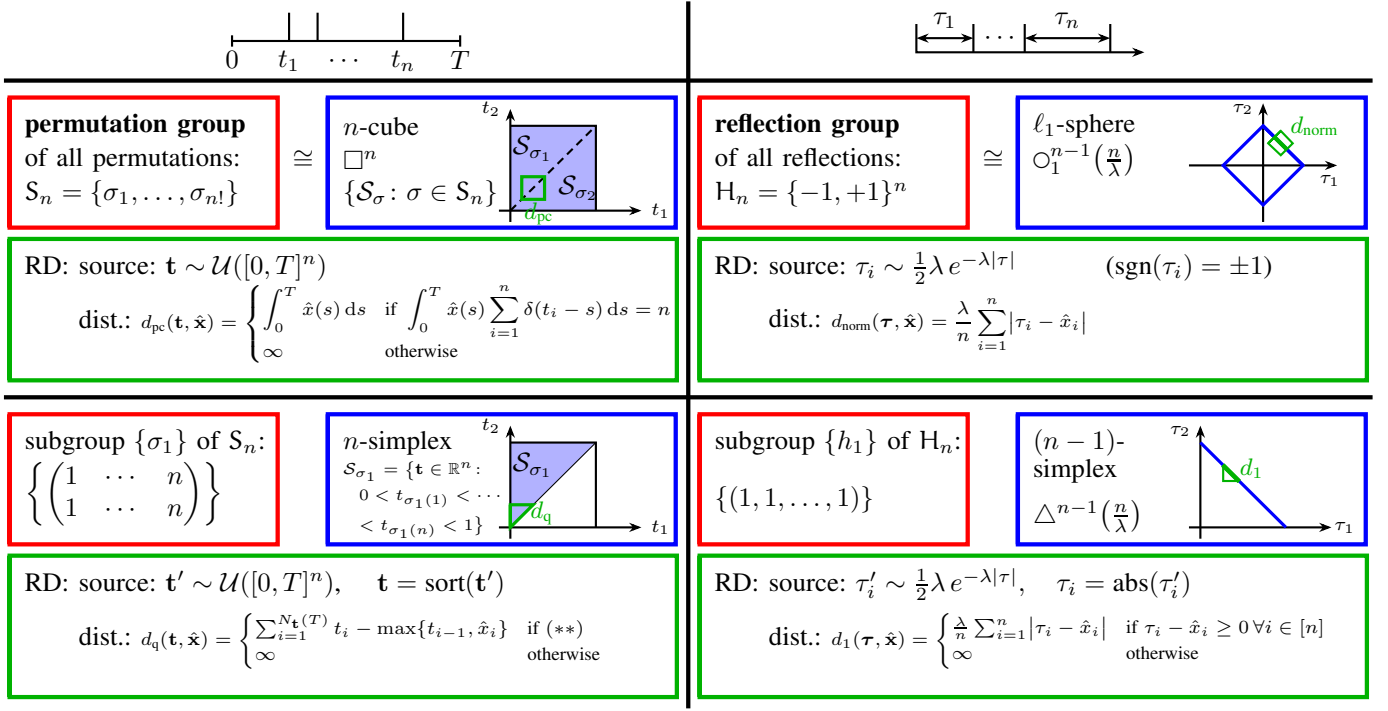

These four different rate-distortion problems are briefly summarized
in Figure~\ref{fig:bigpicture}. There the left column describes the
event description and the right column the inter-event interval
description of the process; and the rows distinguish whether the
complete group or only its trivial subgroup is used as a geometric
description of the source.

Note that a crucial aspect to these geometrical descriptions of the
source and rate-distortion problem is an appropriate choice of the
distortion measure. We introduce here the concept of a \emph{natural
  choice} of distortion measure that guarantees that the distortion
set around a codeword has a similar shape to the source set, leading
to particularly easy formulations of the ball- or sphere-covering
problem and rate-distortion function.

Finally in Section~\ref{sec:hyperhedral-symmetries}, we focus on the
particular choice of permutation and reflection group as our main tool
of geometric description. We show that they both can be derived from
the so-called \emph{hyperoctahedral group}, a group that describes all
symmetries of a hypercube or of a regular hyperoctahedron; or more
precisely, we will give construction of the hyperoctahedral group from
the permutation group and the reflection group via the semidirect
product. Thereby we demonstrate the connections between the
hyperoctahedral group and the symmetries of a Poisson process.

\subsection*{Notation and Definitions}

We use bold font $\vect{x}$ to denote vectors; for sets we use a
calligraphic font $\set{X}$; groups and group elements are denoted by
the Euler font $\const{G}$; and for a graph and its vertex and edge
set, we use $\mat{\Gamma} = (\mat{V}, \mat{E})$.

By $\| \cdot \|_p$ we denote the $\ell_p$-norm. Real metric and normed
(linear) spaces are denoted by $(X, d)$ and $(X, \|\cdot\|)$,
respectively, the former with metric $d$ and the latter with origin
and norm $\|\cdot\|$. Sometimes we do not specify the metric or norm
and only write $X$. In particular, the $n$-dimensional Euclidean space
$(\Reals^n, \|\cdot\|_2)$ is denoted by $E_n$.  We define
\begin{IEEEeqnarray}{c}
  [n] \eqdef \{1,2,\ldots,n\}.
\end{IEEEeqnarray}
For $r>0$,
\begin{IEEEeqnarray}{c}
  \sphere_1^{n-1}(r) \eqdef \left\{\vect{x} \in \Reals^n \colon
    \sum_{i=1}^{n} \abs{x_i} = r \right\}
\end{IEEEeqnarray}
denotes the \emph{$\ell_1$-sphere} of radius $r$. Its first-orthant
(``hyper-surface'') $(n-1)$-simplex is given as
\begin{IEEEeqnarray}{c}
  \simplex^{n-1}(r) \eqdef \left\{\vect{x} \in \Reals^n \colon
    \sum_{i=1}^{n} x_i = r, \; x_i \geq 0 \;\forall i\in [n]
  \right\}.
  \IEEEeqnarraynumspace
\end{IEEEeqnarray}
Furthermore, we define the $n$-dimensional unit-(hyper)cube
$\cube^n \eqdef [0,1]^n$.  We define $\octahedron^n$ to be the
$n$-dimensional regular (hyper)octahedron with vertices
$\{\pm \vect{e}_1, \pm\vect{e}_2, \ldots, \pm \vect{e}_n\}$, where
$\{\vect{e}_i\}_{i\in [n]}$ is an orthonormal basis of $E_n$.

We define the \emph{permutation group} as the set of all permutations
$\sigma_i$ on $n$ objects,
\begin{IEEEeqnarray}{c}
  \const{S}_n \eqdef \{\sigma_1, \ldots, \sigma_{n!}\},
\end{IEEEeqnarray}
with composition ``$\circ$'' as group operation. Its identity element
$\const{e}_{\const{S}_n}=\sigma_1$ is the identity mapping.
\begin{remark}
  We are aware that in the literature this group is usually called
  ``the symmetric group'', while the term ``permutation group'' is
  used for subgroups of $\const{S}_n$. However, in order to make a
  more clear distinction between $\const{S}_n$ and the group of
  symmetries introduced later, we have decided to avoid the name
  ``symmetric group''.
\end{remark}

The \emph{reflection group} $\const{H}_n$ is defined as the $n$-fold
direct product of $\const{H}\eqdef \{+1,-1\}$,
\begin{IEEEeqnarray}{c}
  \const{H}_n \eqdef \{+1,-1\}^n,
\end{IEEEeqnarray}
with regular multiplication ``$\cdot$'' as group operation.

A \emph{group isomorphism} between two groups $(\const{G}, \ast)$ and
$(\const{H},\circ)$ is a bijective mapping
$\varphi\colon \const{G} \to \const{H}$ such that
\begin{IEEEeqnarray}{c}
  \varphi(\const{g} \ast \const{g}')
  = \varphi(\const{g}) \circ \varphi(\const{g}'), \quad \forall
  \const{g}, \const{g}'\in\const{G}.
\end{IEEEeqnarray}
Two groups $\const{G}$ and $\const{H}$ are \emph{isomorphic}, written
as $\const{G} \cong \const{H}$, if there exists a group isomorphism
between them.

By $\vol_n$ we denote the $n$-dimensional Lebesgue measure. The
logarithm $\log(\cdot)$ is to base $2$; $\cl{\cdot}$ denotes the
closure of a set; $\co{\cdot}$ denotes the convex hull of a set of
points; 
and $\I{\textnormal{statement}}$ represents the indicator function,
which equals $1$ if the statement holds true and $0$ otherwise.

\section{Rate Distortion and $\ell_{\infty}$-Ball Covering for the
  Homogeneous Poisson Process}
\label{sec:rate-dist-ball}

\subsection{The Hypercube and the Permutation Group}
\label{sec:permutationgroup}

Each realization of a homogeneous Poisson point process over the
duration $[0, T]$ has some number of points $n$ and can thus be
described by an $n$-tuple $(t_1, t_2, \ldots, t_n)$ where
$t_1 < t_2 < \cdots < t_n$. For convenience, we also define
$t_0 \eqdef 0$.

Considering all permutations of each $n$-tuple and without loss of
generality setting $T=1$, the (closure of the) $n$-tuples and their
permutations form a unit $n$-cube $\cube^{n}$. We will next associate
this cube with a group denoted $\const{G}^{\textnormal{perm}}_n$.

To that goal, for any permutation $\sigma\in\const{S}_n$, we define
the (``hypervolume'') $n$-simplex as
\begin{IEEEeqnarray}{c}
  \set{S}_{\sigma} \eqdef \bigl\{ \vect{t} \in \Reals^{n} \colon 0 <
  t_{\sigma(1)} < t_{\sigma(2)} < \cdots < t_{\sigma(n)} < 1
  \bigr\}.
  \label{eq:S-sigma}
  \IEEEeqnarraynumspace
\end{IEEEeqnarray}
Note that this $n$-simplex $\set{S}_{\sigma}$ ``triangulates'' the
$n$-cube $\cube^n$, and that the closure of the union of all these
$n$-simplices forms the $n$-cube $\cube^n$ (compare also with left
upper quadrant of Figure~\ref{fig:bigpicture}).
\begin{definition}
  \label{def:Gsym}
  We define
  \begin{IEEEeqnarray}{c}
    \const{G}^{\textnormal{perm}}_n
    \eqdef \{ \set{S}_{\sigma} \colon \sigma \in \const{S}_n \}
  \end{IEEEeqnarray}
  to be the associated group of the hypercube, where its group
  operation ``$\ast$'' is defined by means of the group operation
  ``$\circ$'' of $\const{S}_n$:
  \begin{IEEEeqnarray}{c}
    \set{S}_{\sigma_i} \ast \set{S}_{\sigma_j} \eqdef
    \set{S}_{\sigma_i \circ \sigma_j}.
  \end{IEEEeqnarray}
\end{definition}
Note that the collection of ordered $n$-tuples describing a
homogeneous Poisson process is a subgroup of
$\const{G}^{\textnormal{perm}}_n$. Moreover, note that
$\const{G}^{\textnormal{perm}}_n$ is, by definition, isomorphic to
$\const{S}_n$.
\begin{proposition}
  \label{prop:iso-perm}
  The mapping
  $\varphi_{\textnormal{p}} \colon \const{S}_n \to
  \const{G}^{\textnormal{perm}}_n, \; \sigma \mapsto \set{S}_{\sigma}$
  is an isomorphism.
\end{proposition}
\begin{remark}
  Because of this isomorphism we henceforth also refer to
  $\const{G}^{\textnormal{perm}}_n$ as the permutation group.
\end{remark}
From Proposition~\ref{prop:iso-perm} we immediately get the identity
element of $\const{G}^{\textnormal{perm}}_n$:
\begin{IEEEeqnarray}{rCl}
  \const{e}_{\const{G}^{\textnormal{perm}}_n}
  & = & \varphi_{\textnormal{p}}(\const{e}_{\const{S}_n})
  = \set{S}_{\sigma_1}
  \nonumber\\
  & = & \bigl\{ \vect{t} \in \Reals^{n}\colon 0 < t_{1} < t_{2} <
  \cdots < t_{n} < 1 \bigr\}.
  \label{eq:e_G_perm}
\end{IEEEeqnarray}
Thus, the subgroup $\{\const{e}_{\const{G}^{\textnormal{perm}}_n}\}$
describes realizations of a homogeneous Poisson point process with $n$
ordered points over the duration $[0,1]$ (compare also with left lower
quadrant of Figure~\ref{fig:bigpicture}).

\subsection{Rate-Distortion Problem on the Permutation Group}
\label{sec:permutationgroup-ratedistortion}

When considering a rate-distortion problem for a certain source,
sometimes there exists a ``natural choice'' of a distortion measure
that ``preserves'' the geometry of the source. The most typical
example is the $\ell_2$-distortion measure for the Gaussian source,
where the $\ell_2$-distortion ball has the same fundamental
shape\footnote{Recall that for large $n$, with very high probability
  the source output sequences lie in a thin sphere. Thus when
  referring to the ``source shape'' we actually consider the geometry
  of the typical sequences.} as the source ball. Based on such a
geometric picture one can then use the idea of sphere- or
ball-covering to derive (the converse to) the rate-distortion theorem
(see, e.g., \cite{shenmoserpfister21_1}).

In the following we will show how a natural choice of distortion
measure can be found for the permutation group
$\const{G}^{\textnormal{perm}}_n$ ($n$-cube) and for its subgroup
$\{\const{e}_{\const{G}^{\textnormal{perm}}_n}\}$ ($n$-simplex) and
how they lead to two well-known rate-distortion problems of the
homogeneous Poisson process, namely the point-covering distortion
problem \cite{lapidothmalarwang11_1, lapidothmalarwang11_2}
and the canonical queueing distortion problem
\cite{colemankiyavashsubramanian08_1}.  We will refer to these two
cases as $\ell_{\infty}$-ball covering for the homogeneous Poisson
process.\footnote{Other rate-distortion problems for the Poisson
  process that we do not consider here can be found, e.g., in
  \cite{rubin74_1, rubin74_2, rubin73_1, shendewagner19_1}.}

In this work, we distinguish the \emph{source space} $\set{X}$, which
is the set of all possible source output sequences of a given length
$n$, and the \emph{source set} $\set{T}$, which describes the geometry
of the length-$n$ typical sequences. The \emph{codeword space}
$\hat{\set{X}}$ is the set of possible codeword sequences of length
$n$.

In this section, (the closure of) our source set $\set{T}$ is the
$n$-dimensional hypercube $\cube^n$ or the $n$-simplex
$\set{S}_{\sigma_1}$. Note that for the two rate-distortion problems
on the permutation group we have $\set{X} = \set{T}$.

Then, for a given codeword $\vect{x}^* \in \ecl{\hat{\set{X}}}$ and
for an allowed distortion $D$ (normalized by the total duration $T$,
yielding $0<D\leq 1$), we define the \emph{distortion set}
$\set{E}_{\vect{x}^*}(D)$ as follows:
\begin{IEEEeqnarray}{c}
  \set{E}_{\vect{x}^*}(D) \eqdef \lim_{\hvect{x} \rightarrow
    \vect{x}^*}\bigl\{\vect{t} \in \set{X}\colon 
  d(\vect{t}, \hvect{x}) \leq D \bigr\},
  \label{eq:def-distortion-set}
\end{IEEEeqnarray}
where a limit is required here because due to the strict inequalities
in \eqref{eq:S-sigma} the codeword space $\hat{\set{X}}$ is not
closed.

Next we make more precise what we mean by ``a natural choice'' of
distortion measure.
\begin{definition}[Natural Distortion Measure]
  \label{def:matching-sets-E1}
  A distortion measure $d(\cdot,\cdot)$ is said to be \emph{natural}
  if the distortion set defined in \eqref{eq:def-distortion-set}
  preserves the geometry of the corresponding source set $\set{T}$ in
  the sense that there exists a unique
  $\vect{x}^* \in \ecl{\hat{\set{X}}}$ such that
  \begin{IEEEeqnarray}{c}
    \bigcl{\set{E}_{\vect{x}^*}(1)} =
    \bigcl{\co{\{\vect{0}\}\cup\set{T}}}. 
    \label{eq:matching-sets-E1}
  \end{IEEEeqnarray}
\end{definition}
Note that due to the normalization of timings we can set $T=1$ without
loss of generality.

\subsubsection{Point-Covering Distortion}
\label{paragraph:pointcovering}

A rate-distortion codeword for the homogeneous Poisson process for the
point-covering distortion is a $\{0,1\}$-valued signal $\hvect{x}$ on
the interval $[0,1]$ (see \cite{lapidothmalarwang11_1,
  lapidothmalarwang11_2}). The signal $\hvect{x}$ partitions $[0,1]$
into a $1$-valued, Lebesgue-measurable set $\set{A}_{\hvect{x}}$ and a
$0$-valued set $\cset{A}_{\hvect{x}}$.  The \emph{point-covering
  distortion measure} $d_{\textnormal{pc}}(\vect{t}, \hvect{x})$
between a point process realization $\vect{t}$ and a codeword
$\hvect{x}$ is the Lebesgue measure of $\set{A}_{\hvect{x}}$, if
$\set{A}_{\hvect{x}}$ covers $\vect{t}$; and is infinite otherwise:
\begin{IEEEeqnarray}{rCl}
  d_{\textnormal{pc}}(\vect{t},\hvect{x})
  & = &
  \begin{cases}
    \D \int_0^T \hat{x}(s)\dd s
    & \D \text{if } \int_0^T \hat{x}(s)
    \sum_{i=1}^n \delta(t_i-s)\dd s = n,
    \\
    \infty  & \text{otherwise}
  \end{cases}
  \nonumber\\*
\end{IEEEeqnarray}
(see also ``dist.'' in left upper quadrant of
Figure~\ref{fig:bigpicture}).

Let $\vect{t}$ be a Poisson point pattern of $n$ points.  Each
codeword $\hvect{x}$ with $\set{A}_{\hvect{x}}$ of Lebesgue measure
$D$ ($0 < D \leq 1$) gives the distortion set
$\set{E}_{\hvect{x}}(D) \subset\Reals^n$:
\begin{IEEEeqnarray}{rCl}
  \set{E}_{\hvect{x}}(D)
  & = &  \bigl\{ \vect{t} \in \cube^n\colon
  d_{\textnormal{pc}}(\vect{t}, \hvect{x}) = D \bigr\}
  \nonumber\\
  & = &  \bigl\{ \vect{t} \in \cube^n\colon t_i \in
  \set{A}_{\hvect{x}},  \; \forall i \in [n] \bigr\}
  \label{eq:cube-distortionset}
\end{IEEEeqnarray}
for $\hvect{x}$ such that $\vol_1(\set{A}_{\hvect{x}}) = D$.  Clearly,
$\vol_{n}(\set{E}_{\hvect{x}}(D)) = D^n$.  The minimal number of
distortion sets needed to cover the $n$-cube is thus
\begin{IEEEeqnarray}{c}
  \frac{\vol_{n}(\cube^n)}{\vol_{n}(\set{E}_{\hvect{x}}(D))}
  = \frac{1}{D^n}.
\end{IEEEeqnarray}
This gives the minimal rate of $\log(1/D)$ bits per point (i.e., per
dimension).

When again including the duration $T$, we note that for a homogeneous
Poisson process of rate $\lambda$, the expected number of points
$\E{n} = \lambda T$. The resulting minimal average number of bits per
unit time is therefore lower-bounded by
$\frac{\E{n}}{T} \log (1/D) = \lambda \log (1/D)$, which is indeed the
rate-distortion function for the Poisson process with the
point-covering distortion measure \cite{lapidothmalarwang11_1,
  lapidothmalarwang11_2}.

We have shown how the rate-distortion problem of the homogeneous
Poisson process with point-covering distortion can be understood as
covering an $n$-cube with the distortion set in
\eqref{eq:cube-distortionset}.  This cube covering perspective is
similar to the converse proof given in \cite{mazumdarwang12_1,
  lapidothmalarwang11_2}.  The resulting rate-distortion function
shows this simple form in principle because the distortion set in
\eqref{eq:cube-distortionset} is matched to the source set, i.e., in
other words, the point-covering distortion is a natural distortion
measure for $\set{T}=\cube^n$ in that it satisfies the condition given
in Definition~\ref{def:matching-sets-E1}. The geometry of the
distortion set in $\Reals^n$ matches that of the permutation group
$\const{G}^{\textnormal{perm}}_n$.

\subsubsection{Canonical Queueing Distortion}
\label{paragraph:queueing}

In this section, we describe point process realizations of $n$ points
over $[0,T]$ as a tuple $\vect{t}$ of timings such that
$t_1 < t_2 < \cdots < t_n$. Thus, when the timings are normalized by
the duration $T$, we have $\vect{t} \in \set{S}_{\sigma_1}$, and
$\set{T} = \set{S}_{\sigma_1}$ is the source set (see also left lower
quadrant of Figure~\ref{fig:bigpicture}).

For the queueing rate-distortion problem, a codeword $\hvect{x}$ is
also a point process realization over $[0,T]$ with timing description
in the same ordered fashion
$\hat{x}_1 < \hat{x}_2 < \hat{x}_3 < \cdots$.

Let $N_{\vect{P}}(\cdot)$ be the counting function on the point
process $\vect{P}$. The queueing distortion measure is defined as
\cite{colemankiyavashsubramanian08_1}
\begin{IEEEeqnarray}{rCl}
  \IEEEeqnarraymulticol{3}{l}{%
    d_{\textnormal{q}}(\vect{t}, \hvect{x})
  }\nonumber\\*%
  & \eqdef & \left\{ \,
  \begin{IEEEeqnarraybox}[\mystrut][c]{l'l}
    \frac{1}{T}\sum\limits_{i=1}^{N_{\vect{t}}(T)} \bigl( t_i -
    \max\{t_{i-1}, \hat{x}_i\} \bigr)
    & \text{if } N_{\hvect{x}}(T) = N_{\vect{t}}(T)
    \\[-3mm]
    \IEEEeqnarraymulticol{2}{r}{%
      \text{and } N_{\hvect{x}}(s) \ge N_{\vect{t}}(s)
      \;\forall s\in[0,T],}%
    \\[1mm]
    \infty  & \text{otherwise}.
  \end{IEEEeqnarraybox}
  \right. 
  \IEEEeqnarraynumspace
  \label{eq:queueing-distortion}
\end{IEEEeqnarray}
Without loss of generality, we continue this section by considering
normalized timings for point process realizations (timings normalized
by the duration $T$).  The conditions under which
$d_{\textnormal{q}}(\vect{t}, \hvect{x})$ is finite can be rewritten
as follows.
\begin{proposition}
  \label{prop:finite-condition-causal-timings}
  For two (normalized) point process realizations
  $\vect{t}, \hvect{x} \in \set{S}_{\sigma_1}$ with
  $N_{\hvect{x}}(1) = N_{\vect{t}}(1) = n$, the following equivalence
  holds:
  \begin{IEEEeqnarray}{c}
    N_{\hvect{x}}(s) \geq N_{\vect{t}}(s)  \; \forall s \in [0,1]
    \iff  t_i \geq \hat{x}_i \; \forall i\in[n].
    \IEEEeqnarraynumspace
    \label{equivalence-queueing-condition}
  \end{IEEEeqnarray}
\end{proposition}

Following similar arguments as in
Section~\ref{paragraph:pointcovering}, we will proceed to show next
how the rate distortion problem of the homogeneous Poisson process
with a canonical queueing distortion measure can be understood as
covering the subgroup\footnote{We loosely say ``covering a group'',
  but actually it means covering the union of all sets that
  constitutes the group.}
$\{\const{e}_{\const{G}^{\textnormal{perm}}_n}\}$ (a simplex) with a
natural distortion set.

Recall from \eqref{eq:def-distortion-set} that the distortion set
under distortion $D$ for a given $\vect{x}^* \in \ecl{\hat{\set{X}}}$
is
\begin{IEEEeqnarray}{c}
  \set{E}_{\vect{x}^*}(D) \eqdef \lim_{\hvect{x} \rightarrow
    \vect{x}^*}\bigl\{\vect{t} \in \set{S}_{\sigma_1}\colon 
  d_{\textnormal{q}}(\vect{t}, \hvect{x})  \leq D \bigr\}.
  \label{eq:2}
\end{IEEEeqnarray}
In general, the shape of this set is quite complicated because of the
maximum function contained in the queueing distortion measure
\eqref{eq:queueing-distortion}. To help the reader with the following
observations, in Appendix~\ref{appendix:queueing-examples} we present
a more detailed study of this exact shape for the case of $n=2$.

We observe that for $\vect{x}^*=\vect{0}$, we have
$\set{E}_{\vect{0}}(1) = \set{S}_{\sigma_1}$ and therefore, according
to Definition~\ref{def:matching-sets-E1}, the queueing distortion
$d_{\textnormal{q}}(\cdot, \cdot)$ is a natural distortion measure for
the source set $\set{T}=\set{S}_{\sigma_1}$.  Note that for
$\vect{x}^*=\vect{0}$ and arbitrary $0<D\leq 1$
\begin{IEEEeqnarray}{c}
  \set{E}_{\vect{0}}(D) = D \set{S}_{\sigma_1},
\end{IEEEeqnarray}
where $D \set{S}_{\sigma_1}$ denotes $\set{S}_{\sigma_1}$ scaled
linearly by $D$. Thus, in this case the distortion set is a scaled
version of the source set.

On the other hand, for a codeword $\hvect{x} \neq \vect{0}$, the
distortion set $\set{E}_{\hvect{x}}(D)$ is not necessarily a simplex
(see Example~\ref{ex:shape-of-distortion-set} and
Figure~\ref{fig:triangle_queue2}b in
Appendix~\ref{appendix:queueing-examples}).  Nevertheless, $\hvect{x}$
can always be chosen such that the volume of the distortion set is
preserved in the sense that for
$N_{\hvect{x}}(1) = N_{\vect{t}}(1)=n$,
\begin{IEEEeqnarray}{c}
  \sup_{\hvect{x}\in\set{S}_{\sigma_1}}
  \vol_n\bigl(\set{E}_{\hvect{x}}(D)\bigr) 
  = \vol_n\bigl(D \set{S}_{\sigma_1}\bigr).
  \IEEEeqnarraynumspace
  \label{eq:volume-preserve}
\end{IEEEeqnarray}
Therefore, the minimal number of distortion sets needed to cover the
source $n$-simplex $\set{T}= \set{S}_{\sigma_1}$ is
\begin{IEEEeqnarray}{rCl}
  \frac{\vol_{n}(\set{S}_{\sigma_1})}{\sup_{\hvect{x}\in\set{S}_{\sigma_1}} 
    \vol_{n}(\set{E}_{\hvect{x}}(D))}
  & = & \frac{\vol_{n}(\set{S}_{\sigma_1})}{\vol_n(D \set{S}_{\sigma_1})}
  = \left(\frac{1}{D}\right)^n.
  \IEEEeqnarraynumspace
\end{IEEEeqnarray}
This gives again $\log (1/D)$ bits per point (per dimension) and,
following the same arguments as in
Section~\ref{paragraph:pointcovering}, we obtain the minimal number of
bits per unit time $\lambda \log (1/D)$. This corresponds to the
rate-distortion function for the Poisson process with the canonical
queueing distortion measure \cite{colemankiyavashsubramanian08_1}.

\begin{remark}
  Note that when $D$ is very small, for all
  $\vect{t} \in \set{E}_{\hvect{x}}(D)$ the following holds:
  \begin{IEEEeqnarray}{c}
    \hat{x}_{i+1} > t_i \geq \hat{x}_i \; \forall  i\in[n-1],  \text{
      and } 1 > t_n \geq \hat{x}_n,
  \end{IEEEeqnarray}
  and thus
  \begin{IEEEeqnarray}{c}
    d_{\textnormal{q}}(\vect{t}, \hvect{x})
    = \sum_{i=1}^{N_{\vect{t}}(1)} (t_i - \hat{x}_i).
  \end{IEEEeqnarray}
  Therefore and because of
  Proposition~\ref{prop:finite-condition-causal-timings}, we see that
  in this situation
  \begin{IEEEeqnarray}{c}
    \set{E}_{\hvect{x}}(D) = \hvect{x} + \cl{D \delta},
    \label{eq:8}
  \end{IEEEeqnarray}
  where $D\delta$ is a scaled version of the simplex
  \begin{IEEEeqnarray}{c}
    \delta  \eqdef \left\{
      \Delta\vect{t} \in \Reals^{n}\colon \sum_{i=1}^{n}
      \Delta t_i < 1, \; \Delta t_i > 0 \; \forall i \in [n] \right\}.
    \IEEEeqnarraynumspace
    \label{eq:e_tildeG}
  \end{IEEEeqnarray}
  Thus, here the distortion set is shaped like a scaled version of the
  simplex $\delta$. Note that albeit the simplex $\delta$ is not
  similar to the source set $\set{S}_{\sigma_1}$ for $n\ge 3$, they
  have the same $n$-dimensional volume.  For an exposition on $n=2$,
  see Example~\ref{ex:shape-standard} in
  Appendix~\ref{appendix:queueing-examples}, where the distortion set
  $\set{E}_{\hvect{x}}(D)$ is represented by the red triangle in
  Figure~\ref{fig:triangle_queue1}.
\end{remark}

\section{Rate Distortion and $\ell_1$-Sphere Covering for the
  Homogeneous Poisson Process}
\label{sec:rate-dist-sphere}

We have shown in Sections~\ref{paragraph:pointcovering} and
\ref{paragraph:queueing} that with the timing description of
point-process realizations, two known rate-distortion problems for the
homogeneous Poisson point process (namely with point-covering
distortion and with the canonical queueing distortion) can be
understood geometrically as minimal covering problems for the
permutation group (cube) and its subgroup (simplex), respectively.  It
is natural at this point to ask whether other interesting
rate-distortion problems arise by considering minimal coverings of
another group and its subgroup.

To that goal, recall that the inter-event interval $\tau$ of a
homogeneous Poisson point process is exponentially distributed:
\begin{IEEEeqnarray}{c}
  \tau \sim \lambda \ope^{-\lambda \tau} \I{\tau \ge 0}.
  \label{expsource}
\end{IEEEeqnarray}
We now make the following two motivating observations:
\begin{enumerate}
\item If a vector of inter-event intervals describes the realization
  of a Poisson point process according to \eqref{expsource}, then it
  lies close to a simplex $\simplex^{n-1}(n/\lambda)$ in $\Reals^n$ if
  $n$ is large.
\item The number of symmetries of $\simplex^{n-1}(n/\lambda)$ in
  $\Reals^n$ can be increased by reflections, through which the
  simplex becomes the $\ell_1$-sphere $\sphere_1^{n-1}(n/\lambda)$.
\end{enumerate}
Based on these two observations and analogously to what we have shown
for the permutation group in Section~\ref{sec:rate-dist-ball}, in the
rest of this section we study the reflection group and its associated
rate distortion problems, namely the \emph{Laplacian-$\ell_1$} and the
\emph{exponential onesided-$\ell_1$} rate-distortion problems.

\subsection{The Regular Hyperoctahedron and the Reflection Group}
\label{sec:reflectiongroup}

We proceed to show that the $\ell_1$-sphere $\sphere_1^{n-1}(1)$,
i.e., the boundary of a regular hyperoctahedron in $\Reals^n$, is
isomorphic to the reflection group $\const{H}_n$.

We define for any $\vect{r}=(r_1,\ldots,r_n)$ and
$\set{A} \subseteq \Reals^n$,
\begin{IEEEeqnarray}{c}
  \vect{r} \odot \set{A} \eqdef \bigl\{ \vect{x} \in \Reals^n\colon
  x_i = r_i a_i \, \forall i\in[n] \text{ and } \vect{a}\in \set{A} 
  \bigr\}.
  \IEEEeqnarraynumspace
  \label{eq:1}
\end{IEEEeqnarray}
\begin{definition}
  \label{def:Greflect}
  We define
  \begin{IEEEeqnarray}{c}
    \const{G}^{\textnormal{refl}}_n \eqdef \bigl\{\const{h} \odot
    \simplex^{n-1}(1) \colon \const{h} \in \const{H}_n \bigr\}
  \end{IEEEeqnarray}
  with group operation ``$\ast$'' given as follows:
  \begin{IEEEeqnarray}{rCl}
    \IEEEeqnarraymulticol{3}{l}{%
      \bigl(\const{h} \odot \simplex^{n-1}(1) \bigr)
      \ast \bigl(\const{h}'\odot \simplex^{n-1}(1)\bigr)
    }\nonumber\\*\quad%
    & \eqdef & (\const{h} \cdot \const{h}') \odot
    \simplex^{n-1}(1),
    \qquad \const{h},\const{h}'\in\const{H}_n.
    \label{eq:G_reflect-operation}
  \end{IEEEeqnarray}
\end{definition}
We note that, by definition, $\const{G}^{\textnormal{refl}}_n$ is
isomorphic to $\const{H}_n$.
\begin{proposition}
  \label{prop:iso-reflect}
  The mapping
  \begin{IEEEeqnarray}{c}
    \varphi_{\textnormal{r}}\colon \const{H}_n \to
    \const{G}^{\textnormal{refl}}_n, \quad
    \const{h} \mapsto \const{h}\odot\simplex^{n-1}(1)
  \end{IEEEeqnarray}
  is an isomorphism.
\end{proposition}
\begin{remark}
  Because of this isomorphism we henceforth also refer to
  $\const{G}^{\textnormal{refl}}_n$ as the reflection group.
\end{remark}
The identity element of $\const{G}^{\textnormal{refl}}_n$ is
\begin{IEEEeqnarray}{rCl}
  \const{e}_{\const{G}^{\textnormal{refl}}_n}
  = \varphi_{\textnormal{r}}(\const{e}_{\const{H}_n})
  = \simplex^{n-1}(1)
\end{IEEEeqnarray}
with $\const{e}_{\const{H}_n}$ being the identity element of
$\const{H}_n$.

We will show in the following section that the reflection group
$\const{G}^{\textnormal{refl}}_n$ and its subgroup
$\{\const{e}_{\const{G}^{\textnormal{refl}}_n}\}$ with their
respective natural distortion measures yield the
\emph{Laplacian-$\ell_1$} and the \emph{exponential onesided-$\ell_1$}
rate-distortion problem.

\subsection{Rate-Distortion Problem on the Reflection Group}

Similarly to the discussion for the permutation group in
Section~\ref{sec:permutationgroup-ratedistortion}, we now consider the
rate-distortion and minimal covering problem on the reflection group
and its subgroup: $\const{G}^{\textnormal{refl}}_n$ ($\ell_1$-sphere
$\sphere_1^{n-1}(1)$) and
$\{\const{e}_{\const{G}^{\textnormal{refl}}_n}\}$ (simplex
$\simplex^{n-1}(1)$).

Recall from Observation~1) at the start of this section that the
inter-event interval realizations generated by \eqref{expsource} lie
almost surely in the thin shell around $\simplex^{n-1}(n/\lambda)$
(for a sufficiently large number of intervals); compare also with the
schematic in right lower quadrant in
Figure~\ref{fig:bigpicture}. Furthermore, we implement Observation~2)
by labeling each inter-event interval independently with $-1$ or $1$
equiprobably.  This labeling creates a new source
$\tau_{\textnormal{s}}$ of \emph{signed} inter-event intervals that
has a Laplacian distribution:
\begin{IEEEeqnarray}{c}
  \tau_{\textnormal{s}} \sim \frac{\lambda}{2} \ope^{-\lambda
    |\tau_{\textnormal{s}}|}.
  \label{eq:lapsource}
\end{IEEEeqnarray}
Its realizations of length-$n$ sequences lie almost surely in the thin
shell around the $\ell_1$-sphere $\sphere_1^{n-1}(n/\lambda)$ (compare
also with the schematic in right upper quadrant in
Figure~\ref{fig:bigpicture}). We use again the notions of the source
set $\set{T}$ and natural distortion measure introduced in
Section~\ref{sec:permutationgroup-ratedistortion}, and we consider two
source sets\footnote{Note that $\set{X} \neq \set{T}$ for the two
  rate-distortion problems on the reflection group. The source spaces
  for the Laplacian and exponential source are
  $\set{X}=\{\vect{x} \in \Reals^n \colon x_i \neq 0 \, \forall
  i\in[n]\}$ and
  $\set{X}=\{\vect{x} \in \Reals^n \colon x_i > 0 \, \forall
  i\in[n]\}$, respectively.}  $\set{T}=\sphere_1^{n-1}(n/\lambda)$ or
$\set{T}=\simplex^{n-1}(n/\lambda)$, with their respective natural
distortion measures.  We refer to these two cases as $\ell_1$-sphere
covering.

Again, using the same ideas based on the geometric picture of source
set and distortion set, one can derive the rate-distortion functions
for these two rate-distortion problems, see for example
\cite{shenmoserpfister21_1}. In the following we will only briefly
summarize the results and omit their geometric derivations.

\subsubsection{Laplacian-$\ell_1$ Rate-Distortion Problem}
\label{paragraph:ell1}

The \emph{normalized $\ell_1$-distortion measure} is defined as
\begin{IEEEeqnarray}{c}
  d_{\textnormal{norm}}(\vect{x},\hvect{x})
  \eqdef \frac{\lambda}{n} \sum_{i=1}^{n} |x_i- \hat{x}_i|,
  \label{eqdef:normalizedell1}
\end{IEEEeqnarray}
where $\lambda$ is the parameter of the Laplacian source in
\eqref{eq:lapsource}. It is easy to verify that the normalized
$\ell_1$-distortion measure is a natural distortion measure for
$\set{T}=\sphere_1^{n-1}(n/\lambda)$.

The following lemma follows directly from
\cite[Lemma~6]{sikoyluogluvishwanath14_1}.
\begin{lemma}
  \label{lem:lapl1}
  For a Laplacian source \eqref{eq:lapsource} and the normalized
  $\ell_1$-distortion measure \eqref{eqdef:normalizedell1}, the rate
  distortion function is
  \begin{IEEEeqnarray}{c}
    R_{\textnormal{Laplacian}}(D) = \log \left(\frac{1}{D}\right)
    \I{0<D\leq 1}.
    \label{eq:rlapl}
  \end{IEEEeqnarray}
\end{lemma}

\subsubsection{Exponential Onesided-$\ell_1$ Rate-Distortion Problem}
\label{paragraph:onesidedell1}

The \emph{normalized onesided $\ell_1$-distortion measure} is defined
as
\begin{IEEEeqnarray}{c}
  d_1(\vect{x},\hvect{x}) \eqdef
  \begin{cases}
    \frac{\lambda}{n}\sum_{i=1}^{n} |x_i-\hat{x}_i|
    & \textnormal{if } x_i-\hat{x}_i \geq 0 \; \forall i \in [n],
    \\
    \infty & \textnormal{otherwise},
  \end{cases}
  \IEEEeqnarraynumspace
  \label{eqdef:onesidedl1}
\end{IEEEeqnarray}
where $\lambda$ is the parameter of the exponential source in
\eqref{expsource}.  Again, one can verify that $d_1(\cdot,\cdot)$ is a
natural distortion measure for $\set{T}=\simplex^{n-1}(n/\lambda)$.

The following lemma follows directly from
\cite[Lemma~2]{sikoyluogluvishwanath14_1}.
\begin{lemma}
  \label{lem:expl1}
  For an exponential source \eqref{expsource} and the normalized
  onesided $\ell_1$-distortion measure \eqref{eqdef:onesidedl1}, the
  rate-distortion function is
  \begin{IEEEeqnarray}{c}
    \label{eq:rexp}
    R_{\textnormal{Exponential}}(D) =
    \log \left(\frac{1}{D}\right) \I{0<D\leq 1}.
  \end{IEEEeqnarray}
\end{lemma}

Note that when viewing $n$ as the number of points in a point process
realization, the rate-distortion functions in Lemmas~\ref{lem:lapl1}
and \ref{lem:expl1} are the same function, measured in bits per symbol
(per point).  This gives $\log (1/D)$ bits per point just as the
results in Sections~\ref{paragraph:pointcovering} and
\ref{paragraph:queueing}. Therefore, when considering the
rate-distortion problem under the inter-event interval description of
a homogeneous Poisson process of rate $\lambda$, we see that we use
$n\log(1/D)$ bits to describe a complete sequence of (random) duration
$T_{\textnormal{tot}}(n)$, and thus the number of bits per unit time,
for large $n$, is
\begin{IEEEeqnarray}{c}
  \lim_{n\to\infty} \frac{n\log\left(\frac{1}{D}\right)}{
    T_{\textnormal{tot}}(n)} 
  = \lim_{n\to\infty} \frac{\log\left(\frac{1}{D}\right)}{
    \frac{T_{\textnormal{tot}}(n)}{n}} 
  = \frac{\log\left(\frac{1}{D}\right)}{\frac{1}{\lambda}},
\end{IEEEeqnarray}
and we obtain the minimal number of bits per unit time
$\lambda \log (1/D)$.

It is not a coincidence that all four rate-distortion functions in
Sections~\ref{paragraph:pointcovering}, \ref{paragraph:queueing},
\ref{paragraph:ell1}, \ref{paragraph:onesidedell1} are the same. The
reason is that they all have their own natural distortion sets matched
to their source sets, i.e., they all satisfy the criterion given in
Definition~\ref{def:matching-sets-E1}.

To this point, we have presented the rate-distortion problems of the
Poisson process as $\ell_{\infty}$-ball covering in
Section~\ref{sec:rate-dist-ball} and $\ell_1$-sphere covering in
Section~\ref{sec:rate-dist-sphere}. One may wonder why it exactly is
$\ell_{\infty}$ and $\ell_1$.  We attempt to answer this question in
the following section by exploring the \emph{hyperoctahedral group},
which is the group of symmetries of the hypercube or the regular
hyperoctahedron.

\section{The Hyperoctahedral Group and How It Is Generated from the Poisson Process}
\label{sec:hyperhedral-symmetries}

The \emph{hyperoctahedral group}, denoted $\const{O}_n$, describes the
symmetries of both an $n$-dimensional hypercube or an $n$-dimensional
regular hyperoctahedron (see, e.g., \cite{baake84_1}).  In other
words, the $n$-cube and the $n$-dimensional regular hyperoctahedron
have the same group of symmetries and are both realizations of the
group of symmetries $\const{O}_n$ in $\Reals^n$.

This can be understood most easily when realizing that the regular
hyperoctahedron and the hypercube are actually dual (polar) polytopes:
replacing the vertices of one by $(n-1)$-dimensional faces results in
the other and vice-versa. This geometric duality means that we can
inscribe one in the other in such a way that it becomes
straightforward to see that the two share the same symmetries.

In the following we are going to show that the hyperoctahedral group
can be understood as being ``spanned'' (by means of the semidirect
product) by the reflection group $\const{G}^{\textnormal{refl}}_n$ and
the permutation group $\const{G}^{\textnormal{perm}}_n$ (see
Theorem~\ref{thm:Automorphim-On} below). To that goal, we will in a
first step derive the symmetries of a general polytope and describe
them by means of a permutation subgroup over its vertices
(Lemma~\ref{lem:SymP-vertex}). In a second step, we will then relate
the group of symmetries of the regular hyperoctahedron and of the
hypercube with the automorphism group of their respective graph
(Theorems~\ref{thm:Sym-Aut-octahedron} and \ref{thm:Sym-Aut-cube}).

The ultimate goal of this section is to give a (partial) answer to our
original question posed at the end of the previous section: Why
exactly do $\ell_{\infty}$ (cube) and $\ell_1$ (octahedron) show up?
Recall that the Poisson process possesses two geometric descriptions
(sets), namely the two simplices shown on the left and right column in
Figure~\ref{fig:bigpicture}. For both descriptions we ``added'' some
symmetrization, namely permutation (left column in
Figure~\ref{fig:bigpicture}) or reflections (right column in
Figure~\ref{fig:bigpicture}), to obtain the hypercube or the regular
hyperoctahedron, respectively. The choice of these symmetrizations are
not merely ad-hoc. 
To show this, we propose an iterative algorithm that iteratively
``expands'' the pair of source sets $\set{T} = \set{S}_{\sigma_1}$ and
$\set{T} = \simplex^{n-1}$ (and their respective symmetries) until
their group of symmetries become isomorphic (at which point the
algorithm ends). Applying this algorithm to the two simplices in the
lower blue boxes in Figure~\ref{fig:bigpicture}, we arrive at a
hypercube and a regular hyperoctahedron in the end. This is explained
in more detail in Section~\ref{sec:algorithm} and
Figure~\ref{fig:algorithm}.

\subsection{Preliminaries}
\label{sec:preliminaries}

\begin{definition}
  For $\varphi \colon X \to Y$, $x \mapsto \varphi(x)$ and
  $\set{A} \subset X$, $\set{B} \subset Y$, we say
  ``$\varphi(\set{A})=\set{B}$'' to mean ``$\varphi(x) \in \set{B}$
  if, and only if, $x \in \set{A}$.''
\end{definition}
\begin{definition}
  An \emph{isometry} of a metric space $(X,d)$ is a surjective
  function
  \begin{IEEEeqnarray}{c}
    \varphi \colon X \to X, \; x \mapsto \varphi(x)
  \end{IEEEeqnarray}
  such that
  \begin{IEEEeqnarray}{c}
    d(x, x') = d\bigl(\varphi(x),\varphi(x')\bigr),
    \quad \forall x, x' \in X.
  \end{IEEEeqnarray}
  The set\footnote{Note that $\const{Isom}(X)$ can be seen as a group
    where function composition is its group operation.} of all
  isometries of a metric space $X$ is denoted by $\const{Isom}(X)$.
\end{definition}
Thus, an isometry is a mapping that preserves distances. Typical
examples are rotations, reflections, or translations in the Euclidean
space.

These isometries are now the basis for capturing the concept of
\emph{symmetries} of an object.
\begin{definition}[The Group of Symmetries of a Set]
  \label{def:group-symmetries-set}
  For any $\set{A} \subset E_n$, the \emph{group of symmetries} of
  $\set{A}$, denoted $\const{Sym}_{E_n}(\set{A})$, is defined as
  \begin{IEEEeqnarray}{c}
    \const{Sym}_{E_n}(\set{A})
    \eqdef \bigl\{ \varphi \in \const{Isom}(E_n) \colon
    \varphi(\set{A}) = \set{A} \bigr\},
  \end{IEEEeqnarray}
  where the group operation is function composition.
\end{definition}
So, any symmetry of some set is an isometry that maps the set back to
itself.

We also use the following standard group-theoretic definitions for a
\emph{group action on a set} and the \emph{semidirect product}, see
for example \cite{rotman95_1} as a reference.

\begin{definition}[Group Action on a Set]
  \label{def:groupaction}
  The group $\const{G}$ acts on a set $\set{X}$ if there is a function
  \begin{IEEEeqnarray}{c}
    f\colon \const{G} \times \set{X} \to \set{X}, \; (\const{g},
    x) \mapsto \const{g}x
    \label{eq:9}
  \end{IEEEeqnarray}
  satisfying the following conditions:\footnote{Note that in this
    definition and for the rest of the article, we use the
    juxtaposition notation introduced in \eqref{eq:9} for (left) group
    actions.}
  \begin{itemize}
  \item $\const{e}x = x$, $\forall x \in \set{X}$;
  \item $\const{g}_1(\const{g}_2x) = (\const{g}_1\cdot \const{g}_2)x$,
    $\forall \const{g}_1, \const{g}_2 \in \const{G}$, $x \in \set{X}$.
  \end{itemize}
  Here $\const{e} \in \const{G}$ is the identity element and
  ``$\cdot$'' is the group operation of $\const{G}$. We say that
  \emph{$\const{G}$ acts on $\set{X}$ with (left) action\footnote{Note
      that when we use the term ``a group acts on a set'', we always
      refer to the \emph{left} group action if not otherwise
      specified.}~$f$.}
\end{definition}
\begin{definition}[Internal Semidirect Product]
  \label{def:internal-semidirect}
  Let $\const{H}_1$ and $\const{H}_2$ be subgroups of $\const{G}$
  equipped with the group operation ``$\cdot$'' and with the identity
  element $\const{e}_{\const{G}}$. We say that $\const{G}$ is the
  \emph{internal semidirect product} of $\const{H}_1$ by
  $\const{H}_2$, denoted
  $\const{G} = \const{H}_1 \rtimes \const{H}_2$, if
  \begin{itemize}
  \item $\const{H}_1$ is a normal subgroup\footnote{Note that we make
      no assumption regarding $\const{G}$ being Abelian.} of
    $\const{G}$, i.e.,
    $\const{g}\cdot \const{H}_1 = \const{H}_1 \cdot \const{g}$ for all
    $\const{g}\in\const{G}$;
  \item $\const{H}_1 \cap \const{H}_2 = \{\const{e}_{\const{G}}\}$;
  \item $\const{G} = \const{H}_1 \cdot \const{H}_2$.
  \end{itemize}
\end{definition}
Finally, we are going to need \emph{affine maps}:
\begin{definition}
  \label{def:affine-map}
  Let $V$, $W$ be real normed linear spaces. We say a map
  $\alpha \colon V \to W$, $v \mapsto \alpha(v)$ is \emph{affine} if
  \begin{IEEEeqnarray}{rCl}
    \alpha\bigl(s v + (1-s) v'\bigr)
    & = & s\, \alpha(v) + (1-s)\, \alpha(v'), \qquad \nonumber\\*
    && \hfill \forall v, v' \in V, \;
    s \in [0,1].
  \end{IEEEeqnarray}
\end{definition}


\subsection{Polytopes and Group of Symmetries of Polytopes}
\label{sec:polyt-group-symm}

\begin{definition}
  A \emph{polytope} $\set{P}$ is defined as the convex hull of a
  finite, nonempty set of points in $\Reals^n$, for some $n\geq
  2$. The \emph{dimension} of $\set{P}$, denoted $\dim{\set{P}}$, is
  defined as the dimension of the smallest linear subspace containing
  $\set{P}$. We call $\set{P}$ a \emph{$k$-polytope} when
  $\dim{\set{P}}=k$.
\end{definition}

\begin{definition}[Extreme Points and Vertices of a Polytope]
  \label{def:extreme-points-and-vertex}
  For a compact convex set $\set{S} \subset \Reals^n$, we define
  $\vect{x} \in \set{S}$ to be an \emph{extreme point} if, and only
  if, $\set{S} \setminus \{\vect{x}\}$ is also convex. The set of all
  extreme points of $\set{S}$ is denoted $\Extr(\set{S})$. When
  $\set{S}$ is a polytope, $\Extr(\set{S})$ is the set of vertices of
  the polytope.  We denote the set of vertices of a polytope $\set{P}$
  as $\set{V}(\set{P})$.
\end{definition}

So, let $\set{P}$ be an $n$-polytope with its set of vertices
\begin{IEEEeqnarray}{c}
  \set{V}(\set{P}) = \{\vect{x}_1,\vect{x}_2, \ldots, \vect{x}_m\},
  \label{eq:P-n-polytope-m-vertices}
\end{IEEEeqnarray}
where $\vect{x}_i \in \Reals^n$, $\forall i \in [m]$, and without loss
of generality assume that
\begin{IEEEeqnarray}{c}
  \frac{1}{m}\sum_{i \in [m]} \vect{x}_i = \vect{0}.
  \label{eq:6}
\end{IEEEeqnarray}
Let the permutation group $\const{S}_m$ act on $\set{V}(\set{P})$ with
action
\begin{IEEEeqnarray}{c}
  \const{g}\vect{x}_i \eqdef \vect{x}_{\const{g}i}, \quad
  \forall  \const{g} \in \const{S}_m
\end{IEEEeqnarray}
(i.e., the vertices of $\set{P}$ are permuted), and define the
subgroup $\const{G}_{\set{V}(\set{P})}$ of $\const{S}_m$ as
\begin{IEEEeqnarray}{rCl}
  \const{G}_{\set{V}(\set{P})}
  & \eqdef & \bigl\{ \const{g} \in
  \const{S}_m \colon \|\vect{x}_i - \vect{x}_j\|_2 =
  \|\const{g}\vect{x}_i - \const{g}\vect{x}_j\|_2, \, \qquad\qquad
  \nonumber\\*
  && \hfill \forall
  \vect{x}_i,\vect{x}_j \in \set{V}(\set{P}) \bigr\},
  \IEEEeqnarraynumspace
  \label{eq:7}
\end{IEEEeqnarray}
i.e., $\const{G}_{\set{V}(\set{P})}$ contains all those permutations
of the vertices of $\set{P}$ that preserves the pairwise
$\ell_2$-distances between the vertices.

We will show next that $\const{G}_{\set{V}(\set{P})}$, which is
defined using merely the vertices of the polytope, describes the
symmetries of this (full-dimensional) $n$-polytope $\set{P}$ in
$\Reals^n$, i.e., $\const{G}_{\set{V}(\set{P})}$ is isomorphic to the
group of symmetries of $\set{P}$.

\begin{lemma}[Group of Symmetries of a Polytope]
  \label{lem:SymP-vertex}
  Consider the $n$-polytope $\set{P}$ and its group
  $\const{G}_{\set{V}(\set{P})}$ as defined above in
  \eqref{eq:P-n-polytope-m-vertices}--\eqref{eq:7}. Then for any
  $\const{g} \in \const{G}_{\set{V}(\set{P})}$, there exists a unique
  affine map
  \begin{IEEEeqnarray}{c}
    \alpha_{\const{g}} \colon \Reals^n \to \Reals^n, \;
    \vect{x} \mapsto \alpha_{\const{g}}(\vect{x})
  \end{IEEEeqnarray}
  satisfying
  \begin{IEEEeqnarray}{c}
    \alpha_{\const{g}}(\vect{x}_i) = \const{g}\vect{x}_i, \quad
    \forall \vect{x}_i \in \set{V}(\set{P}).
  \end{IEEEeqnarray}
  Moreover,
  \begin{IEEEeqnarray}{c}
    \label{lem-eq-1}
    \const{Sym}_{E_n}(\set{P}) = \bigl\{ \alpha_{\const{g}} \colon
    \const{g} \in \const{G}_{\set{V}(\set{P})} \bigr\},
  \end{IEEEeqnarray}
  and
  \begin{IEEEeqnarray}{c}
    \label{lem-eq-2}
    \const{Sym}_{E_n}(\set{P}) \cong \const{G}_{\set{V}(\set{P})}.
  \end{IEEEeqnarray}
\end{lemma}
\begin{IEEEproof}
  See Appendix~\ref{appendix:lem:SymP-vertex}.
\end{IEEEproof}

\subsection{Graph of Polytopes and Automorphism Group of a Graph}
\label{sec:graphs-polyt-autom}

\begin{definition}[Basic Graph Definitions, see, e.g.,
  \cite{diestel17_1}]
  Let $\mat{V} \neq \emptyset$ be a nonempty set and denote the set of
  all $k$-element subsets of $\mat{V}$ by $[\mat{V}]^k$.  An
  \emph{undirected graph} $\mat{\Gamma}=(\mat{V},\mat{E})$ is a pair
  of sets where $\mat{E} \subseteq [\mat{V}]^2$. Here, $\mat{V}$ is
  called the \emph{vertex set} and $\mat{E}$ is called the \emph{edge
    set} of the graph.  We say $v_1, v_2 \in \mat{V}$ are
  \emph{adjacent (vertices)} if $\{v_1,v_2\} \in \mat{E}$.
\end{definition}

\begin{definition}[The Graph of a Polytope]
  \label{def:graph-of-polytope}
  The \emph{graph of a polytope} $\set{P}$, denoted
  $\mat{\Gamma}_{\set{P}}$, is an undirected graph formed by the
  vertices and the $1$-dimensional faces (edges) of the polytope.
\end{definition}

\begin{definition}[Graph Isomorphism]
  Let $\mat{V},\mat{V}' \neq \emptyset$. Two graphs
  $\mat{\Gamma} = (\mat{V},\mat{E})$,
  $\mat{\Gamma}' = (\mat{V}',\mat{E}')$ are \emph{isomorphic} if there
  exists a bijective function $\varphi \colon \mat{V} \to \mat{V}'$
  such that $\{\varphi(v_1), \varphi(v_2)\} \in \mat{E}'$ if, and only
  if, $\{v_1, v_2\} \in \mat{E}$. We call $\varphi$ a \emph{graph
    isomorphism} from $\mat{\Gamma}$ to $\mat{\Gamma}'$.
\end{definition}
\begin{definition}[Graph Automorphism]
  \label{def:graph-automorphism}
  An \emph{automorphism} of a graph $\mat{\Gamma}=(\mat{V},\mat{E})$
  is a graph isomorphism from $\mat{\Gamma}$ to itself. It thus
  follows that an automorphism is a permutation of the vertex set
  $\mat{V}$ that preserves both the \emph{adjacencies} and the
  \emph{nonadjacencies} of the graph $\mat{\Gamma}$.

  The collection of all automorphisms of a graph $\mat{\Gamma}$ is
  denoted $\const{Aut}(\mat{\Gamma})$, the \emph{automorphism group}
  of $\mat{\Gamma}$ equipped with composition as group operation.
\end{definition}

\subsection{Auxiliary Theorems}
\label{sec:isom-group-symm}

We next show that, for both the hypercube and the regular
hyperoctahedron, the group of symmetries and the automorphism group of
its graph are isomorphic. Note that 
this is nontrivial. The crucial point is to realize that a graph
of a polytope ignores distances, but only describes
adjacencies and nonadjacencies. So, as a simple example, consider in
$\Reals^2$ a square and a rectangle: both have the same graph, but
obviously their group of symmetries are not identical.

We start with the regular hyperoctahedron, which allows for a simpler
proof because the distance between any two vertices can only take one
of two possible values.
\begin{theorem}
  \label{thm:Sym-Aut-octahedron}
  The group of symmetries of a regular hyperoctahedron is isomorphic
  to the automorphism group of its graph:
  \begin{IEEEeqnarray}{c}
    \const{Sym}_{E_n}(\octahedron^n) \cong
    \const{Aut}\bigl(\mat{\Gamma}_{\octahedron^n}\bigr).
  \end{IEEEeqnarray}
\end{theorem}
\begin{IEEEproof}
  Let $\mat{\Gamma}_{\octahedron^n}=(\mat{V},\mat{E})$ be the graph of
  $\octahedron^n$ where $\mat{V}=\set{V}(\octahedron^n)$ and therefore
  $\abs{\mat{V}}=2n$.  For any $\const{g}\in \const{S}_{2n}$, let
  $\varphi_{\const{g}} \colon \mat{V} \to \mat{V}$ be a graph
  isomorphism from $(\mat{V},\mat{E})$ to $(\mat{V}, \mat{E}')$ where
  $\varphi_{\const{g}}(\vect{v}) \eqdef \const{g} \vect{v}$,
  $\vect{v} \in \mat{V}$. Then using
  Definition~\ref{def:graph-automorphism},
  $\varphi_{\const{g}} \in \const{Aut}(\mat{\Gamma}_{\octahedron^n})$
  if, and only if, $\mat{E}=\mat{E}'$, which is equivalent to each
  of the following conditions:
  \begin{enumerate}
  \item Adjacency condition: For any
    $\vect{v}, \vect{v}' \in \mat{V}$,
    \begin{IEEEeqnarray}{c}
      \{\vect{v}, \vect{v}'\} \in \mat{E}
      \iff \{\const{g}\vect{v}, \const{g}\vect{v}'\} \in \mat{E}.
    \end{IEEEeqnarray}
  \item Nonadjacency condition: For any
    $\vect{v}, \vect{v}' \in \mat{V}$,
    \begin{IEEEeqnarray}{c}
      \{\vect{v}, \vect{v}'\} \in [\mat{V}]^2\setminus \mat{E}
      \iff \{\const{g}\vect{v}, \const{g}\vect{v}'\} \in
      [\mat{V}]^2\setminus \mat{E}.
      \IEEEeqnarraynumspace
    \end{IEEEeqnarray}
  \end{enumerate}
  Using the equivalence of conditions 1) and 2) above we can write the
  automorphism group as:
  \begin{IEEEeqnarray}{rCl}
    \IEEEeqnarraymulticol{3}{l}{%
      \const{Aut}(\mat{\Gamma}_{\octahedron^n})
    }\nonumber\\*\,%
    & \cong & \{\const{g} \in \const{S}_{2n} \colon
    \textnormal{condition 1) holds} \}
    \\
    & = &  \{\const{g} \in \const{S}_{2n} \colon \textnormal{condition
      1) and 2) hold} \}
    \\
    & = & \bigl\{\const{g} \in \const{S}_{2n} \colon \|\vect{v}-
    \vect{v}'\|_2 = \|\const{g}\vect{v}- \const{g}\vect{v}' \|_2,
    \,  \forall \vect{v}, \vect{v}' \in \set{V}(\octahedron^n) \bigr\}
    \nonumber\\*
    \label{eq:3}
    \\
    & = & \const{G}_{\set{V}(\octahedron^n)},
  \end{IEEEeqnarray}
  where \eqref{eq:3} holds because for any
  $\vect{v}, \vect{v}' \in \mat{V}$
  \begin{itemize}
  \item  $\{\vect{v}, \vect{v}'\} \in \mat{E}
    \iff \| \vect{v}-\vect{v}'\|_2 = \sqrt{2}$, and
  \item  $\{\vect{v}, \vect{v}'\} \in [\mat{V}]^2 \setminus\mat{E}
    \iff  \| \vect{v}-\vect{v}'\|_2 = 2$.
  \end{itemize}
  Furthermore, applying Lemma~\ref{lem:SymP-vertex} we have
  $\const{Sym}_{E_n}(\octahedron^n) \cong \const{G}_{\set{V}(\octahedron^n)}$,
  and therefore
  $\const{Sym}_{E_n}(\octahedron^n) \cong \const{G}_{\set{V}(\octahedron^n)}
  \cong \const{Aut}(\mat{\Gamma}_{\octahedron^n})$. This concludes the
  proof.
\end{IEEEproof}

For a hypercube, vertices can be at various different distances to
each other, depending on their relative position to each other. The
proof of the isomorphism between the group of symmetries and the
automorphism group is thus a bit more involved and moved to the
appendix.
\begin{theorem}
  \label{thm:Sym-Aut-cube}
  The group of symmetries of a hypercube is isomorphic to the
  automorphism group of its graph:
  \begin{IEEEeqnarray}{c}
    \const{Sym}_{E_n}(\cube^n) \cong
    \const{Aut}\bigl(\mat{\Gamma}_{\cube^n}\bigr).
  \end{IEEEeqnarray}
\end{theorem}
\begin{IEEEproof}
  See Appendix~\ref{appendix:thm:Sym-Aut-cube}.
\end{IEEEproof}

\subsection{The Hyperoctahedral Group and its Connection to the
  Permutation and Reflection Group}
\label{subsection:the-hyperoctahedral-group}

As already mentioned, the hyperoctahedral group $\const{O}_n$
describes the symmetries of an $n$-dimensional hypercube or of an
$n$-dimensional regular hyperoctahedron (cross-polytope).

The \emph{order} $\abs{\const{O}_n}$ of the hyperoctahedral group
$\const{O}_n$ is
\begin{IEEEeqnarray}{c}
  \abs{\const{O}_n} = 2^n \, n!\,.
\end{IEEEeqnarray}
For example in three dimensions, $\abs{\const{O}_3}=8\cdot 6 =
48$. Note that $\const{O}_3$ can also be understood as a composition
of (rigid-body) rotation and mirroring, which gives
$\abs{\const{O}_3} = 2 \cdot 24 = 48$.

As we have seen above, the group of symmetries of a hypercube or a
regular hyperoctahedron is isomorphic to the automorphism group of the
corresponding graph. Thus, it is possible to define the
hyperoctahedral group as the automorphism group of the graph of the
hypercube \cite[Lecture~3]{gregor20_1}.

So, let $\const{Z}_2 \eqdef \{0,1\}$ be a group equipped with modulo-2
addition, and let $\const{Z}_2^n$ be its $n$-fold direct product.  It
is known that $\const{Aut}(\mat{\Gamma}_{\cube^n})$ is isomorphic to
the internal semidirect product of $\const{Z}^n_2$ by the permutation
group $\const{S}_n$ \cite{harary00_1}, \cite[Lecture~3]{gregor20_1}:
\begin{IEEEeqnarray}{c}
  \const{Aut}(\mat{\Gamma}_{\cube^n}) \cong \const{Z}_2^n \rtimes
  \const{S}_n.
\end{IEEEeqnarray}
Clearly, $\const{Z}_2^n \cong \const{H}_n$. And since by
Propositions~\ref{prop:iso-perm} and \ref{prop:iso-reflect} we have
$\const{S}_n \cong \const{G}^{\textnormal{perm}}_n$ and
$\const{H}_n \cong \const{G}^{\textnormal{refl}}_n$, we obtain the
following result.
\begin{theorem}
  \label{thm:Automorphim-On}
  \begin{IEEEeqnarray}{c}
    \const{O}_n 
    \cong  \const{G}^{\textnormal{refl}}_n \rtimes
    \const{G}^{\textnormal{perm}}_n.
  \end{IEEEeqnarray}
\end{theorem}

By the third condition in Definition~\ref{def:internal-semidirect} for
the internal semidirect product, we can understand
Theorem~\ref{thm:Automorphim-On} intuitively as the construction of
$\const{O}_n$ by two of its subgroups: the reflection subgroup
$\const{G}^{\textnormal{refl}}_n$ and the permutation subgroup
$\const{G}^{\textnormal{perm}}_n$.


\begin{remark}
  Note that by Frucht's theorem \cite{frucht39_1}, one can construct a
  graph whose automorphism group is isomorphic to
  $\const{O}_n$. Unfortunately, when following Frucht's construction,
  we do not obtain the graph of a hypercube or a hyperoctahedron. This
  is why we had to use Lemma~\ref{lem:SymP-vertex} and the two
  Theorems~\ref{thm:Sym-Aut-octahedron} and \ref{thm:Sym-Aut-cube} to
  formally establish the \emph{hyperoctahedral} group as both the
  symmetries of a polytope and the automorphism group of the graph of
  the polytope.
\end{remark}

\begin{figure*}[htp]
  \centering
  \begin{tikzpicture}
    [scale=0.75,>={Stealth[scale=1]},
    box/.style={rectangle,draw,thick,inner sep=0pt,minimum height=1.5cm,
      minimum width=2cm,align=center}]
    

    \draw[red,ultra thick,->,>={Stealth[scale=1.2]}] (-5,2.1)--(-5,0.5);
    \draw[very thick, densely dashed,<-,>={Triangle[scale=1.2]}] (-4.5,1.3)--(4.5,1.3);
    \node at (0.7,1.7) {$\const{Sym}_{E_n}(\simplex^{n-1})\cong \const{S}_n$};
    \node at (0.7,0.9) {permutations};
    \node[right] at (-8,1.3) {\color{red}$\bigcup_{\const{g}\in\const{S}_n}
      \const{g}\const{S}_{\sigma_1}$}; 
    
    \draw[red,ultra thick,->,>={Stealth[scale=1.2]}] (5.2,-0.5)--(5.2,-2.1);
    \draw[very thick, densely dashed,->,>={Triangle[scale=1.2]}] (-3,-1.3)--(4.7,-1.3);
    \node at (0.7,-0.9) {$\const{Sym}_{E_n}(\cube^{n})\cong \const{O}_n$};
    \node at (0.7,-1.7) {permutations \& reflections};
    \node at (7.4,-1.0) {\color{red}$\bigcup_{\const{g}\in\const{O}_n}
      \const{g}\simplex^{n-1}$}; 
    \node at (7.8,-1.8) {\color{red}$=\bigcup_{\const{g}\in\const{H}_n}
      \const{g}\simplex^{n-1}$}; 
    
    \begin{scope}[xshift=5mm]
      \draw[thick] (-8,6.5) to
      node[below,at start] {$0$}
      node[below,at end] {$T$}
      (-4,6.5);
      \draw[thick] (-8,6.4)--(-8,6.6);
      \draw[thick] (-4,6.4)--(-4,6.6);
      \draw[thick] (-7,6.5) to
      node[at start,below] {$t_1$}
      (-7,7);
      \draw[thick] (-6.5,6.5) to (-6.5,7);
      \node at (-6,6.15) {$\cdots$};
      \draw[thick] (-5,6.5) to
      node[at start,below] {$t_n$}
      (-5,7);
    \end{scope}
    
    \begin{scope}[xshift=12cm,yshift=-2mm]
      \draw[thick,->] (-8,7) -- (-8,6.5) -- (-4,6.5);
      \draw[thick] (-7,6.5) -- (-7,7);
      \draw[thick,<->] (-8,6.75) to
        node[midway,above] {$\tau_1$}
        (-7,6.75);
      \node at (-6.5,6.75) {$\cdots$};
      \draw[thick] (-6.1,6.5) -- (-6.1,7);
      \draw[thick,<->] (-6.1,6.75) to
        node[midway,above] {$\tau_n$}
        (-4.6,6.75);
      \draw[thick] (-4.6,6.5) -- (-4.6,7);
    \end{scope}

    \begin{scope}[yshift=-2.2cm,xshift=-1cm,scale=1.4]
      \draw[fill,blue!30] (-3.8,3.5)--(-3.8,5)--(-2.3,5);
      \draw (-3.8,3.5)--(-2.3,5);
      \draw[thick,->] (-4,3.5) to
        node[at end,right] {$t_1$}
        (-1.5,3.5);
      \draw[thick,->] (-3.8,3.3) to
        node[at end,left] {$t_2$}
        (-3.8,5.3);
      \draw[thick] (-3.8,5)--(-2.3,5)--(-2.3,3.5);
      \node at (-3.4,4.6) {$\set{S}_{\sigma_1}$};
    \end{scope}

    \begin{scope}[yshift=-7.2cm,xshift=-1cm,scale=1.4]
      \draw[fill,blue!30] (-3.8,5)--(-2.3,5)--(-2.3,3.5)--(-3.8,3.5);
      \draw[thick,->] (-4,3.5) to
        node[at end,right] {$t_1$}
        (-1.5,3.5);
      \draw[thick,->] (-3.8,3.3) to
        node[at end,left] {$t_2$}
        (-3.8,5.3);
      \draw[thick] (-3.8,5)--(-2.3,5)--(-2.3,3.5);
      \draw[thick,densely dashed] (-3.8,3.5)--(-2.3,5);
      \node at (-3.3,4.6) {$\cube^{n}$};  
    \end{scope}

    \begin{scope}[yshift=-3.2cm,xshift=9.35cm,scale=1.1]
      \draw[thick,->] (-4,3.5) to
        node[pos=0.95,right] {$\tau_1$}
        (-1.5,3.5);
      \draw[thick,->] (-3.8,3.3) to
        node[pos=0.95,left] {$\tau_2$}
        (-3.8,5.6);
      \draw[ultra thick,blue] (-3.8,5)--(-2.3,3.5);
      \node at (-2.7,4.7) {$\simplex^{n-1}$};  
    \end{scope}

    \begin{scope}[yshift=-13.4cm,xshift=10.6cm,scale=2]
      \draw[thick,->] (-4,4.3) to
        node[pos=0.95,below] {$\tau_1$}
        (-1.4,4.3);
      \draw[thick,->] (-2.7,3.3) to
        node[pos=0.95,left] {$\tau_2$}
        (-2.7,5.4);
      \draw[ultra thick,blue] (-2.7,5)--(-2,4.3)
        --(-2.7,3.6)--(-3.4,4.3)--(-2.7,5)--(-2,4.3);
      \node at (-1.9,4.7) {$\sphere_1^{n-1}$};  
    \end{scope}
    
  \end{tikzpicture}
  \caption{Recursive algorithm to create the hyperoctahedral group
    from the symmetries in the two source sets of a Poisson
    process. The symmetries of the $(n-1)$-simplex $\simplex^{n-1}$
    are ``added'' to the $n$-simplex $\set{S}_{\sigma_1}$, resulting
    in the cube $\cube^n$. Then the symmetries of the cube $\cube^n$
    are ``added'' to the $(n-1)$-simplex $\simplex^{n-1}$, resulting
    in the $\ell_1$-sphere $\sphere_1^{n-1}$. The group of symmetries
    of $\cube^n$ and $\sphere_1^{n-1}$ are both isomorphic to
    $\const{O}_n$, and thus the algorithm stops.}
  \label{fig:algorithm}
\end{figure*}
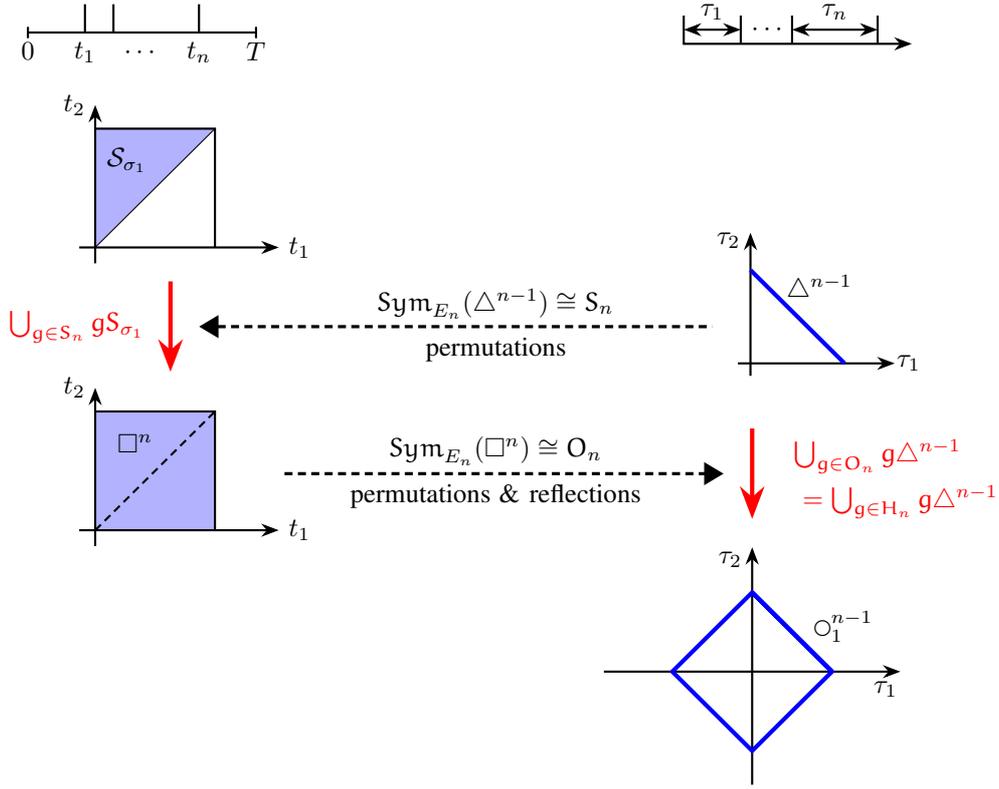

\subsection{From the Poisson Process to $\const{O}_n$: an Algorithm}
\label{sec:algorithm}

We propose the following recursive algorithm.  We are given two
(source) sets $\set{T}^{(a)}_1$ and $\set{T}^{(b)}_1$ and their
corresponding group of symmetries $\const{G}^{(a)}_1$ and
$\const{G}^{(b)}_1$, and we choose some group action according to
Definition~\ref{def:groupaction}.

We now start with the group $\const{G}^{(b)}_1$ acting on
$\set{T}^{(a)}_1$ (according to the chosen group action) to create an
enlarged set $\set{T}^{(a)}_2$. Let $\const{G}^{(a)}_2$ denote the
group of symmetries of $\set{T}^{(a)}_2$.  In a next step we let the
group $\const{G}^{(a)}_2$ act on $\set{T}^{(b)}_1$, and we obtain an
enlarged set $\set{T}^{(b)}_2$ with its group of symmetries
$\const{G}^{(b)}_2$.

We repeat this process until the group of symmetries of the two sets
become isomorphic.
The goal is to choose the group actions in such a way that in each
step $\const{G}^{(j)}_{i}$ is isomorphic to a subgroup of
$\const{G}^{(j)}_{i+1}$, $j\in\{a,b\}$.

Applied to our situation of a hypercube and a regular hyperoctahedron,
we define the group action on $\Reals^n$,
$\const{G} \times \Reals^n \rightarrow \Reals^n, \, (\const{g},
\vect{x}) \mapsto \const{g}\vect{x} $ for
$\const{G} = \const{S}_n, \const{H}_n, \const{O}_n$, respectively,
with matrix multiplications $\forall \vect{x} \in \Reals^n$ as
follows.
\begin{itemize}
\item[(a)] $\const{G} = \const{S}_n; \, \sigma \in \const{S}_n$ is a
  permutation:
  \begin{IEEEeqnarray}{c}
    \const{\sigma}\vect{x} \eqdef \vect{A}_{\const{\sigma}} \vect{x},
  \end{IEEEeqnarray}
  where 
  \begin{IEEEeqnarray}{c}
    [\vect{A}_{\sigma}]_{ij} = \I{j = \sigma(i)};
  \end{IEEEeqnarray}
\item[(b)] $\const{G} = \const{H}_n; \, \const{h} \in \const{H}_n$ is
  a reflection:
  \begin{IEEEeqnarray}{c}
    \const{\const{h}}\vect{x} \eqdef \vect{A}_{\const{h}} \vect{x},
  \end{IEEEeqnarray}
  where 
  \begin{IEEEeqnarray}{c}
    [\vect{A}_{\const{h}}]_{ij} = \const{h}(i) \I{i = j};
  \end{IEEEeqnarray}
\item[(c)] $\const{G} = \const{O}_n; \, \const{o} \in \const{O}_n$ is
  a signed permutation:
  \begin{IEEEeqnarray}{c}
    \const{\const{o}}\vect{x} \eqdef \vect{A}_{\const{o}} \vect{x},
  \end{IEEEeqnarray}
  where
  \begin{IEEEeqnarray}{c}
    [\vect{A}_{\const{o}}]_{ij} = \sgn{\const{o}(i)} \I{j =
      \abs{\const{o}(i)}}. 
  \end{IEEEeqnarray}
\end{itemize}
Here, for $\const{g} \in \const{G}$, $\vect{A}_{\const{g}}$ denotes a
matrix in $\Reals^{n \times n}$; $[\,\cdot\,]_{ij}$ denotes the entry
at the $i$th row and $j$th column of a matrix; and $\abs{\,\cdot\,}$
denotes the absolute value.

We further define for $\set{S} \subset \Reals^n$ and
$\const{g} \in \const{G}$,
\begin{IEEEeqnarray}{c}
  \const{g}\set{S}
  \eqdef \{ \const{g}\vect{x} \colon \vect{x} \in \set{S} \}.
  \label{eq:group-action-gS}
\end{IEEEeqnarray}

Using these groups actions (a)--(c) above and
\eqref{eq:group-action-gS}, we now apply the proposed algorithm to the
source sets\footnote{Recall that, these two source sets arise from the
  timing and interval description of the Poisson process,
  respectively.}  $\set{T}^{(a)}_1 = \set{S}_{\sigma_1}$ and
$\set{T}^{(b)}_1 = \simplex^{n-1}$.
In this case the algorithm stops already after only two steps. The
resulting sets are $\set{T}^{(a)}_2 = \cube^n$ and
$\set{T}^{(b)}_2 = \sphere_1^{n-1}$, both with group of symmetries
isomorphic to $\const{O}_n$.

Figure~\ref{fig:algorithm} depicts a summary of this process.

This shows how our choice of permutation, reflection and
hyperoctahedral group actually arise in a principled way by applying
this algorithm to the source sets of the Poisson process.

Referring back to the graphical summary of
Sections~\ref{sec:rate-dist-ball} and \ref{sec:rate-dist-sphere} in
Figure~\ref{fig:bigpicture}, we conclude from this section that the
hyperoctahedral group unifies the two columns of
Figure~\ref{fig:bigpicture}, demonstrating the symmetries of a Poisson
process.

\section{Discussion}
\label{sec:summary}

A homogeneous Poisson process can be described by event (point)
timings or inter-event (inter-point) intervals (compare with the left
and right columns in Figure~\ref{fig:bigpicture}). Both descriptions
give rise to a group theoretic view point (conditioned on a given
number of points), namely the timing description corresponds to the
\emph{permutation group} and the interval description leads to the
\emph{reflection group}. These in combination with properly chosen
distortion measures allow the corresponding rate-distortion problem to
be expressed as a ball- or sphere-covering problem.

Concretely, in Section~\ref{sec:rate-dist-ball} we considered the
permutation group and its subgroup to describe the point-covering
rate-distortion problem and the queueing rate-distortion problem,
respectively, and showed them to correspond to $\ell_{\infty}$-ball
covering. In Section~\ref{sec:rate-dist-sphere} we considered the
reflection group and its subgroup to describe the exponential onesided
$\ell_1$-rate-distortion problem and the Laplacian
$\ell_1$-rate-distortion problem, respectively, and showed them to
correspond to $\ell_1$-sphere covering.

We also defined the \emph{natural distortion measure} which guarantees
the distortion set around a codeword has a similar shape to the source
set. And in Section~\ref{sec:hyperhedral-symmetries}, we presented the
hyperoctahedral group which can be realized as a hypercube or a
regular hyperoctahedron, and we showed that the permutation group and
the reflection group give a construction of the hyperoctahedral group
via the semidirect product. This demonstrates the connections between
the hyperoctahedral group and the symmetries of a Poisson process.

We also would like to point out that the Poisson point process induces
asymptotically a uniform distribution over each of the source-set
simplices (see illustration in lower left and lower right blue boxes
in Figure~\ref{fig:bigpicture}). This is one of the reasons why the
ratio of volume of source set to volume of distortion set eventually
leads to the required description rate. Although similar
sphere-covering arguments apply even if we did not have a uniform
distribution over the source set, the converse based on sphere
covering with equally-sized distortion balls will not be tight
anymore, as the compression could be improved by smaller distortion
balls in areas of higher probability.

Our geometric approach also works in the case of an inhomogeneous
Poisson process for the situation of point covering
\cite{lapidothmalarwang11_2}. There we need to rescale time in the
following way: for a fixed (infinitesimal small) interval
$[t,t+\dd t]$, we define a rate
$\tilde{\lambda} \eqdef \frac{\lambda(t) \dd t}{T}$ and assume a given
distortion $\tilde{D}$. Then, the geometric arguments from
Section~\ref{paragraph:pointcovering} yield the lower bound
$\lambda(t) \dd t \log(1/\tilde{D})$ to the minimum number of bits
over the interval $[t,t+\dd t]$. Integrating all these lower bounds
over $t$, where $\tilde{D}$ is replaced by $D(t)$, and minimizing over
the choice of $D(t)$ then yields the rate-distortion function as given
in \cite[Th.~2]{lapidothmalarwang11_2}.

\appendices

\section{Distortion Sets for the Queueing Distortion}
\label{appendix:queueing-examples}

In this section, we elaborate some arguments in
Section~\ref{paragraph:queueing} about the (shape of the) queueing
distortion set, with two illustrative examples for
$N_{\hvect{x}}(1) = N_{\vect{t}}(1)=2$.

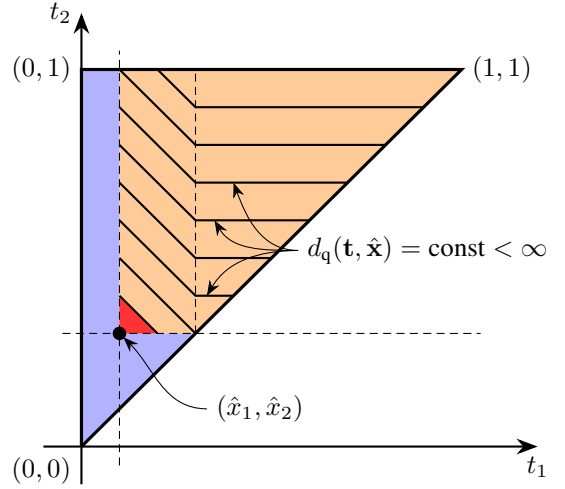
\begin{figure}[htbp]
  \centering
  \begin{tikzpicture}
    [scale=0.5,>={Stealth[scale=1.2]},
    dot/.style={circle,draw=black,fill=black,thick,
      inner sep=0pt,minimum size=1.5mm}]

    \draw[thick,->] (-1,0) to
      node [at end, below] {$t_1$}
      (12,0);
    \draw[thick,->] (0,-1) to
      node [at end, left] {$t_2$}
      (0,11.5);

    \fill[color=orange!40] (1,3)--(3,3)--(10,10)--(1,10)--(1,3);
    \fill[color=blue!30] (1,3)--(3,3)--(0,0)--(0,10)--(1,10)--(1,3);
    \fill[color=red!80] (1,3)--(2,3)--(1,4)--(1,3);
    \draw[very thick] (0,0) to
      node [at end, right] {$(1,1)$}
      (10,10) to
      node [at end, left] {$(0,1)$}
      (0,10) to
      node [at end, below left] {$(0,0)$}
      (0,0);
    \node[dot] at (1,3) {};
    \draw[<-] (1.1,2.9) to [out=-45,in=180]
      node [at end,right] {$(\hat{x}_1,\hat{x}_2)$}
      (3.3,1);
    \draw[densely dashed] (1,-0.5)--(1,10.5);
    \draw[densely dashed] (-0.5,3)--(10.5,3);
    \draw[densely dashed] (3,3)--(3,10);
    \draw[thick] (1,4)--(2,3);
    \draw[thick] (1,5)--(3,3);
    \draw[thick] (1,6)--(3,4)--(4,4);
    \draw[thick] (1,7)--(3,5)--(5,5);
    \draw[thick] (1,8)--(3,6)--(6,6);
    \draw[thick] (1,9)--(3,7)--(7,7);
    \draw[thick] (1,10)--(3,8)--(8,8);
    \draw[thick] (2,10)--(3,9)--(9,9);
    \draw[<-] (3.5,6) to [out=-60,in=180]
      node [at end,right] {$d_{\textnormal{q}}(\vect{t},\hvect{x}) =
        \textnormal{const} < \infty$}
      (5.7,5.2);
    \draw[<-] (4,7) to [out=-70,in=180] (5.7,5.2);
    \draw[<-] (3.3,4) to [out=60,in=180] (5.7,5.2);
  \end{tikzpicture}
  \caption{The $2$-simplex illustrated here represents the source set
    $\set{T} = \set{S}_{\sigma_1}$ for $n=2$.  It is partitioned into
    four regions by a given $\hvect{x}=(\hat{x}_1, \hat{x}_2)$.  The
    orange partition $\set{R}_{d_{\textnormal{q}}<\infty}(\hvect{x})$
    consists of all realizations $\vect{t}=(t_1, t_2)$ where the
    queueing distortion $d_{\textnormal{q}}(\vect{t}, \hvect{x})$ is
    finite.  On the other hand, for all $\vect{t}$ in the blue
    partitions, $d_{\textnormal{q}}(\vect{t}, \hvect{x}) = \infty$.
    The black contour lines depict ``isolines'' of constant
    distortion. The red triangle depicts the distortion set according
    to a situation described in Example~\ref{ex:shape-standard}.}
  \label{fig:triangle_queue1}
\end{figure}

Recall that for normalized timings (all timings $t_k$ normalized by
duration $T$), $\{\const{e}_{\const{G}^{\textnormal{perm}}_n}\}$ gives
the source set $\set{T}$ for the canonical queueing distortion, i.e.,
$\set{T} = \set{S}_{\sigma_1}$.  For $n=2$, the closure of $\set{T}$
is illustrated in Figure~\ref{fig:triangle_queue1} as the $2$-simplex
with its set of vertices $\set{T}^{0}=\{(0,0),\, (0,1), \,(1,1)
\}$. Thus, for a given $\hvect{x}$, the subset
$\set{R}_{d_{\textnormal{q}}<\infty}(\hvect{x}) \subseteq \set{T}$ of
points resulting in a finite queueing distortion
$d_{\textnormal{q}}(\vect{t}, \hvect{x}) < \infty$ can be written as
follows:
\begin{IEEEeqnarray}{rCl}
  \set{R}_{d_{\textnormal{q}}<\infty}(\hvect{x})
  & \eqdef & \bigl\{\vect{t} \in \set{T} \colon
  d_{\textnormal{q}}(\vect{t}, \hvect{x}) < \infty \bigr\}
  \nonumber\\
  & = & \bigl\{\vect{t} \in \set{T}\colon t_k \geq \hat{x}_k \;
  \forall  k\in[n] \bigr\}
  \label{eq:def-R}
\end{IEEEeqnarray}
(compare with Proposition~\ref{prop:finite-condition-causal-timings}).

Figure~\ref{fig:triangle_queue1} illustrates
$\set{R}_{d_{\textnormal{q}}<\infty} (\hvect{x})$ in orange for
$n=2$. With the geometric picture of Figure~\ref{fig:triangle_queue1},
we can now illustrate the distortion set for
$\hvect{x}=(\hat{x}_1, \hat{x}_2)$ in $\Reals^2$ under two disjoint
conditions, see Example~\ref{ex:shape-standard} and
\ref{ex:shape-of-distortion-set}.

\begin{example}
  \label{ex:shape-standard}
  Consider $\hvect{x}=(\hat{x}_1, \hat{x}_2)$ where
  $0 <\hat{x}_1 < \hat{x}_2<1$. Then for $0<D\leq\hat{x}_2-\hat{x}_1$,
  the distortion set is
  \begin{IEEEeqnarray}{ll}
    \set{E}_{\hvect{x}}(D)
    = \Biggl\{ \vect{t} \in \Reals^{2}\colon &  \sum_{i=1}^{2}(t_i -
    \hat{x}_i) \leq D, \nonumber\\
    & t_k \geq \hat{x}_k  \; \forall k \in [2],  t_1 \neq t_2 
    \Biggr\}.
  \end{IEEEeqnarray}
  Figure~\ref{fig:triangle_queue1} shows an exemplary such set in red.
  We see that in this case the distortion set is $\hvect{x} + \cl{D \delta} \setminus \{(\hat{x}_2, \hat{x}_2)\}$ (compare also with \eqref{eq:8}).
\end{example}

\begin{example}
  \label{ex:shape-of-distortion-set}
  Consider $\hvect{x}=(\hat{x}_1, \hat{x}_2)$ where
  $0 <\hat{x}_1 < \hat{x}_2<1$. Then for
  $\hat{x}_2-\hat{x}_1 < D \leq 1$, to obtain the distortion set we
  partition $\set{R}_{d_{\textnormal{q}}<\infty}(\hvect{x})$ (see
  \eqref{eq:def-R}) along
  \begin{IEEEeqnarray}{c}
    \set{L}_2 \eqdef \bigl\{\vect{t}\in\Reals^2\colon t_1=\hat{x}_2
    \bigr\}
  \end{IEEEeqnarray}
  into $\set{R}_1$ and $\set{R}_2$ as follows:
  \begin{IEEEeqnarray}{rCl}
    \set{R}_1
    & \eqdef & \set{R}_{d_{\textnormal{q}}<\infty}(\hvect{x}) \cap
    \set{L}_2^{-}  \nonumber\\
    & = &  \bigl\{(t_1,t_2) \in \set{R}_{d_{\textnormal{q}}<\infty}
    (\hvect{x})\colon \hat{x}_1 \leq t_1 < \hat{x}_2 \leq t_2 \bigr\},
    \IEEEeqnarraynumspace
    \\
    \set{R}_2 & \eqdef & \set{R}_{d_{\textnormal{q}}<\infty}(\hvect{x}) \cap
    \set{L}_2^{+,0}
    \nonumber\\
    & = & \bigl\{(t_1,t_2) \in
    \set{R}_{d_{\textnormal{q}}<\infty}(\hvect{x})\colon
    \hat{x}_1 < \hat{x}_2 \leq t_1 < t_2 \bigr\},
  \end{IEEEeqnarray}
  where
  \begin{IEEEeqnarray}{rCl}
    \set{L}_2^- & \eqdef & \bigl\{\vect{t} \in \Reals^2\colon t_1
    <\hat{x}_2 \bigr\},
    \\
    \set{L}_2^{+,0} & \eqdef & \bigl\{\vect{t} \in \Reals^2\colon t_1
    \geq \hat{x}_2 \bigr\}.
  \end{IEEEeqnarray}
  In Figure~\ref{fig:triangle_queue2}a, $\set{R}_1$ and $\set{R}_2$
  are represented by the light orange and dark orange regions,
  respectively. The types of point pattern corresponding to either
  $\set{R}_1$ and $\set{R}_2$ are illustrated as either the light or
  dark orange spike pattern in the middle of
  Figure~\ref{fig:triangle_queue2}.
  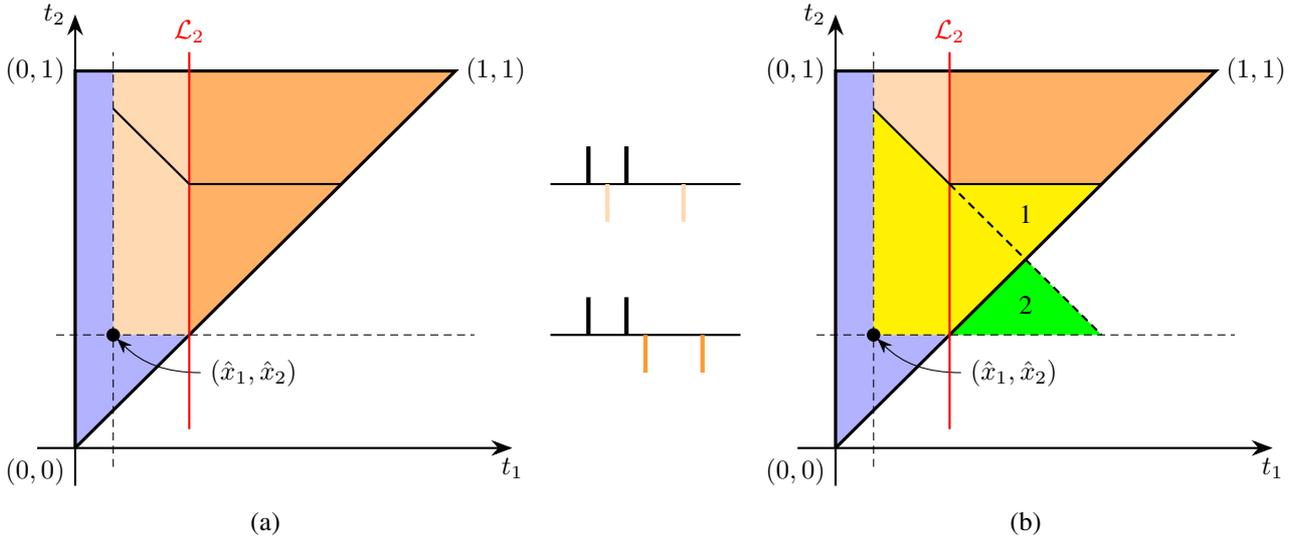
\begin{figure*}[htbp]
    \centering
    \begin{tikzpicture}
      [scale=0.5,>={Stealth[scale=1.2]},
      dot/.style={circle,draw=black,fill=black,thick,
        inner sep=0pt,minimum size=1.5mm}]
      
      \draw[thick,->] (-1,0) to
        node [at end, below] {$t_1$}
        (11.5,0);
      \draw[thick,->] (0,-1) to
        node [at end, left] {$t_2$}
        (0,11.5);
      
      \fill[color=blue!30] (1,3)--(3,3)--(0,0)--(0,10)--(1,10)--(1,3);
      \fill[color=orange!30] (1,3)--(3,3)--(3,10)--(1,10)--(1,3);
      \fill[color=orange!60] (3,3)--(10,10)--(3,10)--(3,3);
      \draw[very thick] (0,0) to
        node [at end, right] {$(1,1)$}
        (10,10) to
        node [at end, left] {$(0,1)$}
        (0,10) to
        node [at end, below left] {$(0,0)$}
        (0,0);
      \node[dot] at (1,3) {};
      \draw[<-] (1.1,2.9) to [out=-45,in=180]
        node [at end,right] {$(\hat{x}_1,\hat{x}_2)$}
        (3.3,2);
      \draw[densely dashed] (1,-0.5)--(1,10.5);
      \draw[densely dashed] (-0.5,3)--(10.5,3);
      \draw[thick,color=red] (3,0.5) to
        node [at end, above] {$\set{L}_2$}
        (3,10.5);
      \draw[thick] (1,9)--(3,7)--(7,7);
      \node at (5,-2) {(a)};
      
      \begin{scope}[xshift=0.5cm]
        \draw[thick] (12,7)--(17,7);
        \draw[ultra thick] (13,7)--(13,8);
        \draw[ultra thick] (14,7)--(14,8);
        \draw[ultra thick,color=orange!30] (13.5,7)--(13.5,6);
        \draw[ultra thick,color=orange!30] (15.5,7)--(15.5,6);
        \draw[thick] (12,3)--(17,3);
        \draw[ultra thick] (13,3)--(13,4);
        \draw[ultra thick] (14,3)--(14,4);
        \draw[ultra thick,color=orange!80] (14.5,3)--(14.5,2);
        \draw[ultra thick,color=orange!80] (16,3)--(16,2);
      \end{scope}

      \begin{scope}[xshift=20cm]
        \draw[thick,->] (-1,0) to
          node [at end, below] {$t_1$}
          (11.5,0);
        \draw[thick,->] (0,-1) to
          node [at end, left] {$t_2$}
          (0,11.5);
      
        \fill[color=blue!30] (1,3)--(3,3)--(0,0)--(0,10)--(1,10)--(1,3);
        \fill[color=orange!30] (1,3)--(3,3)--(3,10)--(1,10)--(1,3);
        \fill[color=orange!60] (3,3)--(10,10)--(3,10)--(3,3);
        \fill[color=yellow] (1,9)--(3,7)--(7,7)--(3,3)--(1,3)--(1,9);
        \fill[color=green] (3,3)--(7,3)--(5,5)--(3,3);
        
        \draw[very thick] (0,0) to
        node [at end, right] {$(1,1)$}
        (10,10) to
        node [at end, left] {$(0,1)$}
        (0,10) to
        node [at end, below left] {$(0,0)$}
        (0,0);
        \node[dot] at (1,3) {};
        \draw[<-] (1.1,2.9) to [out=-45,in=180]
        node [at end,right] {$(\hat{x}_1,\hat{x}_2)$}
        (3.3,2);
        \draw[densely dashed] (1,-0.5)--(1,10.5);
        \draw[densely dashed] (-0.5,3)--(10.5,3);
        \draw[thick,color=red] (3,0.5) to
        node [at end, above] {$\set{L}_2$}
        (3,10.5);
        \draw[thick] (1,9)--(3,7)--(7,7);
        \draw[thick, densely dashed] (3,7)--(7,3);
        \node at (5,6.2) {1};
        \node at (5,3.8) {2};
        \node at (5,-2) {(b)};
      \end{scope}
    \end{tikzpicture}
    \caption{In (a) the region where the distortion measure is finite,
      i.e., $\set{R}_{d_{\textnormal{q}}<\infty}(\hvect{x})$, is divided
      into the light orange region $\set{R}_1$ and the dark orange
      region $\set{R}_2$ by the vertical line
      $\set{L}_2 \eqdef \{\vect{t}\in \Reals^2\colon t_1 =
      \hat{x}_2\}$. The black line depicts a contour line of constant
      distortion.  The middle inset illustrates two different patterns
      of realizations of $(t_1,t_2)$ from $\set{R}_1$ and $\set{R}_2$
      in their respective color.  The point pattern for
      $(\hat{x}_1, \hat{x}_2)$ is illustrated in black. In (b) the
      distortion set $\set{E}_{\hvect{x}}(D)$ as determined by the
      inequalities in \eqref{eq:distortion-r1-r2} is shown in
      yellow. It has the same volume as $D \set{S}_{\sigma_1}$ since
      replacing triangle~2 with triangle~1 preserves the volume.}
    \label{fig:triangle_queue2}
  \end{figure*}
  
  The distortion measure given by \eqref{eq:queueing-distortion} can
  now be rewritten depending on whether the realization
  $\vect{t}=(t_1,t_2)$ lies in $\set{R}_1$ or $\set{R}_2$ as follows:
  \begin{IEEEeqnarray}{c}
    d_{\textnormal{q}}(\vect{t}, \hvect{x}) =
    \begin{cases}
      (t_1 - \hat{x}_1) + (t_2 - \hat{x}_2) 
      & \vect{t} \in \set{R}_1,
      \\
      t_2 - \hat{x}_1 
      & \vect{t} \in \set{R}_2.
    \end{cases}
    \IEEEeqnarraynumspace
    \label{eq:distortion-r1-r2}
  \end{IEEEeqnarray}
  Using \eqref{eq:distortion-r1-r2}, a contour line of constant distortion $D$ is depicted in black in
  Figure~\ref{fig:triangle_queue2}a.  The distortion set $\set{E}_{\hvect{x}}(D)$ is thus as shown in Figure~\ref{fig:triangle_queue2}b in yellow.  So
  clearly the shape is not a simplex. But since the volume of triangle
  $1$ equals that of triangle $2$ in
  Figure~\ref{fig:triangle_queue2}b, the volume of the distortion set
  $\set{E}_{\hvect{x}}(D)$ equals the volume of the simplex
  $D \delta$. In other words, Example~\ref{ex:shape-of-distortion-set}
  gives a distortion set that is not similar (in shape) to the scaled
  source set $D \set{S}_{\sigma_1}$, yet preserves the volume of it.
\end{example}

\section{Proof of Lemma~\ref{lem:SymP-vertex}}
\label{appendix:lem:SymP-vertex}

\subsection{Preliminaries for the Proof}

\begin{theorem}[Mazur-Ulam Theorem (1932) \cite{mazurulam32_1,
    lax02_1}]
  \label{thm:mazur-ulam}
  Any bijective isometry between real normed spaces $X$ and $Y$,
  $\varphi \colon X \to Y$, is an affine map.
\end{theorem}

Applying this theorem to an isometry from $E_n$ to $E_n$
leads immediately to the following corollary.
\begin{corollary}
  \label{coro:isometry-affine}
  Any isometry of a Euclidean space $E_n$ is an affine map.
\end{corollary}

\begin{lemma}[{\cite[Proposition~9.7.1, restated]{berger87_1}}]
  \label{lem:isometry-existence}
  For two sets $\{a_i\}_{i\in[k]}, \{b_i\}_{i\in[k]} \subset E_n$,
  $k \in \Naturals$, where
  \begin{IEEEeqnarray}{c}
    \|a_i-a_j\|_2 = \|b_i-b_j\|_2, \quad \forall i,j \in [k],
  \end{IEEEeqnarray}
  there exists an isometry $\varphi \in \const{Isom}(E_n)$ such that
  $\varphi(a_i) = b_i$.
\end{lemma}

\begin{theorem}[{Krein-Milman Theorem \cite[Theorem~11.6.8,
    restated]{berger87_1}}]
  \label{thm:krein-milman}
  For a compact convex set $\set{S} \subset \Reals^n$,
  \begin{IEEEeqnarray}{c}
    \set{S} = \bigco{\Extr(\set{S})}.
  \end{IEEEeqnarray}
\end{theorem}
\begin{corollary}
  \label{coro:non-extreme-points}
  Let $\set{P} \subset \Reals^n$ be a polytope with the set of
  vertices $\set{V}=\{\vect{v}_i\}_{i\in[m]}$. Then
  $\vect{x} \in \set{P} \setminus \set{V}$ if, and only if, there
  exist $\{t_i\}_{i\in[m]}$, satisfying
  \begin{IEEEeqnarray}{rCl}
    \subnumberinglabel{eq:4}
    0 \leq t_i & < & 1, \quad \forall i\in[m],
    \\
    \sum_{i\in [m]} t_i & = & 1,
  \end{IEEEeqnarray}
  such that
  \begin{IEEEeqnarray}{c}
    \vect{x} = \sum_{i \in [m]} t_i \vect{v}_i.
    \label{eq:5}
  \end{IEEEeqnarray}
\end{corollary}
\begin{IEEEproof}
  This follows directly from
  Definition~\ref{def:extreme-points-and-vertex} and
  Theorem~\ref{thm:krein-milman} (Krein-Milman Theorem).
\end{IEEEproof}
\begin{proposition}
  \label{prop:vertex-to-vertex}
  Let $\set{P}$ be a polytope in $\Reals^n$ with $\set{V}$ being its
  set of vertices. Then for any
  $\alpha \in \const{Sym}_{E_n}(\set{P})$,
  \begin{IEEEeqnarray}{c}
    \alpha(\set{V}) =\set{V}.
  \end{IEEEeqnarray}
\end{proposition}
\begin{IEEEproof}
  Let $\set{V}= \{\vect{v}_i\}_{i\in [m]}$. From
  Corollary~\ref{coro:non-extreme-points} we know that any
  $\vect{x} \in \set{P} \setminus \set{V}$ can be written as in
  \eqref{eq:5} for some $\{t_i\}_{i\in [m]}$ satisfying \eqref{eq:4}.
  And from Corollary~\ref{coro:isometry-affine} and
  Definition~\ref{def:group-symmetries-set} we know that any
  $\alpha \in \const{Sym}_{E_n}(\set{P})$ is an affine map. Combining
  this we obtain
  \begin{IEEEeqnarray}{c}
    \label{eq:alpha-x-non-vertex}
    \alpha(\vect{x}) = \sum_{i \in [m]} t_i \, \alpha(\vect{v}_i), 
  \end{IEEEeqnarray}
  proving that $\alpha(\vect{x}) \in \set{P} \setminus \set{V}$. Thus,
  we have shown that for any $\alpha \in \const{Sym}_{E_n}(\set{P})$,
  \begin{IEEEeqnarray}{c}
    \Bigl( \vect{x} \in \set{P} \setminus \set{V} \Bigr) \implies
    \Bigl( \alpha(\vect{x}) \in \set{P} \setminus \set{V} \Bigr).
    \label{eq:not-vertex}
  \end{IEEEeqnarray}
  By implication this then means that for $\vect{y} \in \set{P}$,
  $\alpha(\vect{y}) \in \set{V}$ only if $\vect{y} \in \set{V}$.

  To show that $\alpha(\set{V})= \set{V}$, we are thus only left to
  show that $\alpha(\vect{y}) \in \set{V}$ if $\vect{y} \in \set{V}$,
  which we will prove by contradiction. To proceed, first note that
  any isometry of $E_n$ is bijective, and therefore the inverse map
  $\alpha^{-1}$ exists and
  $\alpha^{-1} \in \const{Sym}_{E_n}(\set{P})$.  Assume there exists
  $\vect{y} \in \set{V}$ such that
  $\alpha(\vect{y}) \in \set{P} \setminus \set{V}$. This then yields
  that for
  $\vect{x} = \alpha(\vect{y}) \in \set{P} \setminus \set{V}$,
  $\alpha^{-1}(\vect{x}) = \vect{y} \in \set{V}$, which contradicts
  \eqref{eq:not-vertex}. Thus
  $\vect{y} \in \set{V} \implies \alpha(\vect{y}) \in \set{V}$,
  concluding the proof.
\end{IEEEproof}

\subsection{Proof of Lemma~\ref{lem:SymP-vertex}}

Applying Lemma~\ref{lem:isometry-existence}, we know that for any
$\const{g} \in \const{G}_{\set{V}(\set{P})}$ there exists a (not
necessarily unique) isometry $\alpha_{\const{g}} \colon E_n \to E_n$
satisfying $\alpha_{\const{g}}(\vect{x}_i)=\const{g}\vect{x}_i$,
$\forall \vect{x}_i \in \set{V}(\set{P})$. Moreover, from
Corollary~\ref{coro:isometry-affine} we know that this isometry
$\alpha_{\const{g}}$ is an affine map. 
In the following we make use of Definition~\ref{def:affine-map} and
the fact that $\set{P}$ is full-dimensional ($\dim \set{P} = n$) in
the embedding space $\Reals^n$ to show that $\alpha_{\const{g}}$ is
indeed the \emph{unique} affine map (and also the unique isometry)
satisfying $\alpha_{\const{g}}(\vect{x}_i)=\const{g}\vect{x}_i$,
$\forall \vect{x}_i \in \set{V}(\set{P})$.  This is done by
determining the map for
$\vect{x} \in \Reals^n \setminus \set{V}(\set{P})$ by looking at the
following two disjoint cases sequentially:
\begin{enumerate}
\item For $\vect{x} \in \set{P}\setminus \set{V}(\set{P})$, we apply
  Corollary~\ref{coro:non-extreme-points} and
  Definition~\ref{def:affine-map} and obtain
  \begin{IEEEeqnarray}{c}
    \label{eq:case-1}
    \alpha_{\const{g}}(\vect{x})
    = \alpha_{\const{g}}\left( \sum_{i\in [m]} t_i
      \vect{x}_i\right)
    = \sum_{i \in [m]} t_i \alpha_{\const{g}}(\vect{x}_i)
    \IEEEeqnarraynumspace
  \end{IEEEeqnarray}
  with $\{t_i\}_{i\in[m]}$ satisfying \eqref{eq:4}.\footnote{Note that
  	$\alpha_{\const{g}}$ is well defined even though there may exist
  	$\{t'_i\}_{i\in [m]} \neq \{t_i\}_{i\in[m]}$ such that
  	$\vect{x} = \sum_{i \in [m]} t'_i \vect{x}_i$.}
  
\item For $\vect{x} \in \Reals^n\setminus \set{P}$ and because
  $\dim\set{P}=n$, there exists a unique
  $\vect{y} \in \partial\set{P}$ (boundary of $\set{P}$) and some
  scalar $r_{\vect{x}}>1$ given by
  \begin{IEEEeqnarray}{c}
    r_{\vect{x}} \eqdef \inf \{\lambda>0 \colon \vect{x} \in \lambda
    \set{P}\}
  \end{IEEEeqnarray}
  such that $\vect{x}=r_{\vect{x}}\, \vect{y}$.  Since
  $\vect{y} \in \set{P}$, $\alpha_{\const{g}}(\vect{y})$ is already
  well defined and due to Definition~\ref{def:affine-map}, we can thus
  write
  \begin{IEEEeqnarray}{rCl}
    \alpha_{\const{g}}(\vect{y})
    & = &  \alpha_{\const{g}}\bigl((1-1/r_{\vect{x}}) \vect{0} +
    \vect{x}/r_{\vect{x}}\bigr)
    \\
    & = &  (1-1/r_{\vect{x}})\, \alpha_{\const{g}}(\vect{0})+
    \alpha_{\const{g}}(\vect{x})/r_{\vect{x}}
    \IEEEeqnarraynumspace
    \\
    & = & \alpha_{\const{g}}(\vect{x})/r_{\vect{x}},
    \label{line-1}
    \\
    \implies \alpha_{\const{g}}(\vect{x})
    & = &  r_{\vect{x}} \alpha_{\const{g}}(\vect{y}),
    \label{line-2}
  \end{IEEEeqnarray}
  where \eqref{line-1} holds because by \eqref{eq:6}
  \begin{IEEEeqnarray}{c}
    \alpha_{\const{g}}(\vect{0})
    = \alpha_{\const{g}} \left( \frac{1}{m}
      \sum_{i \in [m]} \vect{x}_i \right)
    = \frac{1}{m} \sum_{i \in [m]} \alpha_{\const{g}}(\vect{x}_i)
    = \vect{0}.
    \IEEEeqnarraynumspace
  \end{IEEEeqnarray}
\end{enumerate}

It only remains to justify \eqref{lem-eq-1} and \eqref{lem-eq-2}.

To show \eqref{lem-eq-1}, we first note that we have already shown
that
\begin{IEEEeqnarray}{c}
  \label{eq:f}
  f\colon \const{G}_{\set{V}(\set{P})} \to \const{Isom}(E_n), \;
  \const{g} \mapsto \alpha_{\const{g}}
\end{IEEEeqnarray}
is well defined.  Clearly $f$ is one-to-one, and moreover
$\alpha_{\const{g}}(\set{P}) = \set{P}$,
$\forall \const{g} \in \const{G}_{\set{V}(\set{P})}$. Thus, from
Definition~\ref{def:group-symmetries-set}, we obtain
\begin{IEEEeqnarray}{c}
  \label{line-3}
  \bigl\{ \alpha_{\const{g}} \colon \const{g} \in
  \const{G}_{\set{V}(\set{P})} \bigr\}
  \subseteq \const{Sym}_{E_n}(\set{P}).
\end{IEEEeqnarray}
On the other hand, applying Proposition~\ref{prop:vertex-to-vertex} we
get
\begin{IEEEeqnarray}{rCl}
  \const{Sym}_{E_n}(\set{P})
  & \subseteq & \bigl\{\alpha \in \const{Isom}(E_n)\colon
  \alpha\bigl(\set{V}(\set{P})\bigr) = \set{V}(\set{P}) \bigr\}
  \IEEEeqnarraynumspace
  \\
  & = & \bigl\{ \alpha_{\const{g}} \colon \const{g} \in
  \const{G}_{\set{V}(\set{P})} \bigr\}.
  \label{line-4}
\end{IEEEeqnarray}
Combining \eqref{line-3} and \eqref{line-4} we conclude that
\begin{IEEEeqnarray}{c}
  \label{line-5}
  \const{Sym}_{E_n}(\set{P}) = \bigl\{ \alpha_{\const{g}} \colon
  \const{g} \in \const{G}_{\set{V}(\set{P})} \bigr\}.
\end{IEEEeqnarray}

To show \eqref{lem-eq-2}, we first rewrite \eqref{eq:f} to be a
bijection
\begin{IEEEeqnarray}{c}
  \label{eq:f-bijection}
  f \colon \const{G}_{\set{V}(\set{P})} \to
  \const{Sym}_{E_n}(\set{P}), \;
  \const{g} \mapsto \alpha_{\const{g}}.
\end{IEEEeqnarray}
Thus we are left to show that $f$ is a homomorphism, i.e., that for
any $\const{g}', \const{g} \in \const{G}_{\set{V}(\set{P})}$,
\begin{IEEEeqnarray}{c}
  f(\const{g}'\cdot \const{g})
  = f(\const{g}') \circ f(\const{g}),
  \IEEEeqnarraynumspace
  \label{eq:f-homomorphism-0}
\end{IEEEeqnarray}
where `$\cdot$' and `$\circ$' are the group operations of
$\const{G}_{\set{V}(\set{P})}$ and $\const{Sym}_{E_n}(\set{P})$,
respectively. To this goal, since \eqref{eq:f-homomorphism-0} is
equivalent to
\begin{IEEEeqnarray}{c}
  \alpha_{\const{g}'\cdot \const{g}}
  = \alpha_{\const{g}'}\circ \alpha_{\const{g}},
  \IEEEeqnarraynumspace
  \label{eq:f-homomorphism}
\end{IEEEeqnarray}
in the following we prove that \eqref{eq:f-homomorphism} holds for all
$\vect{x} \in \Reals^n$ by considering three disjoint cases
sequentially:
\begin{enumerate}
\item[(i)] $\vect{x}_i \in \set{V}(\set{P})$:
  \begin{IEEEeqnarray}{rCl}
    \alpha_{\const{g}'\cdot \const{g}}(\vect{x}_i)
    & = & (\const{g}'\cdot \const{g}) \,\vect{x}_i
    =  \const{g}'(\const{g}\vect{x}_i)
    \nonumber\\
    & = & \alpha_{\const{g}'}\bigl(\alpha_{\const{g}}(\vect{x}_i)\bigr)
    = \bigl(\alpha_{\const{g}'} \circ
    \alpha_{\const{g}}\bigr)(\vect{x}_i).
    \IEEEeqnarraynumspace
  \end{IEEEeqnarray}
  
\item[(ii)] $\vect{x} \in \set{P} \setminus \set{V}(\set{P})$: Using
  \eqref{eq:case-1} and (i):
  \begin{IEEEeqnarray}{rCl"s}
    \alpha_{\const{g}'\cdot \const{g}}(\vect{x})
    & = & \alpha_{\const{g}'\cdot \const{g}}\left( \sum_{i \in [m]}
      t_i \vect{x}_i   \right)
    \\
    & = & \sum_{i \in [m]} t_i \, \alpha_{\const{g}'\cdot
      \const{g}}(\vect{x}_i) 
    & (by \eqref{eq:case-1})
    \IEEEeqnarraynumspace
    \\
    & = & \sum_{i \in [m]} t_i \bigl( \alpha_{\const{g}'} \circ
    \alpha_{\const{g}}\bigr) (\vect{x}_i) & (by (i))
    \\
    & = & \bigl(\alpha_{\const{g}'} \circ \alpha_{\const{g}}\bigr)
    \left( \sum_{i \in [m]} t_i \vect{x}_i \right)
    \\
    & = & \bigl(\alpha_{\const{g}'} \circ
    \alpha_{\const{g}}\bigr)(\vect{x}).
  \end{IEEEeqnarray}
  
\item[(iii)] $\vect{x} \in \Reals^n \setminus \set{P}$: Using
  \eqref{line-2} and $\vect{x}=r_{\vect{x}} \vect{y}$ where
  $\vect{y} \in \partial \set{P}$, and noting that from (i) and (ii)
  we have
  $\alpha_{\const{g}'\cdot \const{g}}(\vect{y}) =
  \bigl(\alpha_{\const{g}'} \circ \alpha_{\const{g}}\bigr)(\vect{y})$,
  we obtain
  \begin{IEEEeqnarray}{rCl}
    \alpha_{\const{g}'\cdot \const{g}}(\vect{x})
    & = & r_{\vect{x}} \alpha_{\const{g}'\cdot \const{g}} (\vect{y})
    \nonumber\\
    & = & r_{\vect{x}} \bigl(\alpha_{\const{g}'} \circ
    \alpha_{\const{g}}\bigr)(\vect{y})
    = \bigl(\alpha_{\const{g}'} \circ
    \alpha_{\const{g}}\bigr)(\vect{x}).
    \IEEEeqnarraynumspace
  \end{IEEEeqnarray}
\end{enumerate}
From the Cases (i)--(iii) we conclude that
$\alpha_{\const{g}'\cdot \const{g}} (\vect{x}) =
\bigl(\alpha_{\const{g}'}\circ \alpha_{\const{g}}\bigr)(\vect{x})$,
$\forall \vect{x} \in \Reals^n$, and that therefore
\eqref{eq:f-homomorphism} holds.  In combination with the bijective
map $f$ in \eqref{eq:f-bijection} this means that $f$ is an
isomorphism and thus
$\const{Sym}_{E_n}(\set{P}) \cong \const{G}_{\set{V}(\set{P})}$.

\section{Proof of Theorem~\ref{thm:Sym-Aut-cube}}
\label{appendix:thm:Sym-Aut-cube}

\subsection{Preliminaries for the Proof}

\begin{definition}[Path and Path Length]
  A \emph{path} $\mat{p}=v_0v_1\cdots v_k$ in a graph
  $\mat{\Gamma}=(\mat{V},\mat{E})$ is a sequence of distinct vertices
  where $v_i \in \mat{V}$ and $\{v_{i-1}, v_i\} \in \mat{E}$,
  $\forall i \in [k]$. We say that $\mat{p}$ is a path between $v_0$
  and $v_k$.

  The \emph{length} of a path $\mat{p}$ is the number of edges it
  consists of, and we denote it as $|\mat{p}|=k$. Two distinct
  vertices $v$ and $v'$ are \emph{linked} if there is a path between
  them; and a path linking $v$ and $v'$ with the minimal length is
  called a \emph{shortest path} between $v$ and $v'$.

  For two linked vertices $v \neq v'$ in graph $\mat{\Gamma}$, the
  length of the shortest path between them is denoted by
  $l_{\mat{\Gamma}}(v,v')$.
\end{definition}

\begin{proposition}
  \label{prop:shortest-path-equivalence-distance}
  For $v, v', w, w' \in \set{V}(\cube^n)$, $v \neq v'$, $w \neq w'$,
  \begin{IEEEeqnarray}{rCl}
    \|v-v'\|_2 & = & \|w-w'\|_2
    \nonumber\\
    \iff l_{\mat{\Gamma}_{\cube^n}}(v,v')
    & = & l_{\mat{\Gamma}_{\cube^n}}(w,w').
  \end{IEEEeqnarray}
\end{proposition}

Note that any two distinct vertices of $\mat{\Gamma}_{\cube^n}$ are
linked, and therefore the length of the shortest path is always well
defined here.

\begin{IEEEproof}
  This follows immediately from the fact that the Hamming distance
  between two vertices of $\cube^n$ is equal to $k$ if, and only if,
  their $\ell_2$-distance is $\sqrt{k}$.
\end{IEEEproof}

\begin{lemma}
  \label{lem:all-distance-type-preserving}
  For any $\varphi \in \const{Aut}(\mat{\Gamma}_{\cube^n})$,
  $v, v' \in \set{V}(\cube^n)$,
  \begin{IEEEeqnarray}{c}
    \|v-v'\|_2 = \bigl\|\varphi(v)-\varphi(v')\bigr\|_2.
  \end{IEEEeqnarray}
\end{lemma}
\begin{IEEEproof}
  For $v=v'$, the proposition clearly holds.  For any $v\neq v'$,
  first note that because
  $\varphi \in \const{Aut}(\mat{\Gamma}_{\cube^n})$, any length-$k$
  path $\mat{p}=v v_1 \cdots v_{k-1}v'$ between $v$ and $v'$ in
  $\mat{\Gamma}_{\cube^n}$ can be bijectively mapped to the length-$k$
  path
  $\tilde{\mat{p}}=\varphi(v)\varphi(v_1)\cdots
  \varphi(v_{k-1})\varphi(v')$ between $\varphi(v)$ and $\varphi(v')$
  in $\mat{\Gamma}_{\cube^n}$. This holds for any $k$ and thus we
  have:
  \begin{IEEEeqnarray}{c}
    \label{eq:shortest-path-length-preserving}
    l_{\mat{\Gamma}_{\cube^n}}(v,v')
    = l_{\mat{\Gamma}_{\cube^n}}\bigl(\varphi(v),\varphi(v')\bigr).
  \end{IEEEeqnarray}
  Using \eqref{eq:shortest-path-length-preserving} and
  Proposition~\ref{prop:shortest-path-equivalence-distance} we get
  \begin{IEEEeqnarray}{c}
    \|v-v'\|_2=\|\varphi(v)-\varphi(v')\|_2,
  \end{IEEEeqnarray}
  which concludes the proof.
\end{IEEEproof}

\begin{corollary}
  \label{coro:proof-thm-Sym-Aut-cube}
  Let $\mat{\Gamma}_{\cube^n}=(\mat{V},\mat{E})$. Then for a bijection
  $\varphi: \mat{V} \to \mat{V}, v \mapsto \varphi(v)$, the following
  holds: $\varphi \in \const{Aut}(\mat{\Gamma}_{\cube^n})$ if, and
  only if,
  \begin{IEEEeqnarray}{c}
    \label{eq:only-if}
    \|v-v'\|_2 = \bigl\|\varphi(v)-\varphi(v')\bigr\|_2,
    \quad \forall v, v' \in \mat{V}.
  \end{IEEEeqnarray}
\end{corollary}
\begin{IEEEproof}
  The only-if part follows directly from
  Lemma~\ref{lem:all-distance-type-preserving}. To show that
  \eqref{eq:only-if} implies
  $\varphi \in \const{Aut}(\mat{\Gamma}_{\cube^n})$, we first note
  that from the definition of $\mat{\Gamma}_{\cube^n}$ we have
  \begin{IEEEeqnarray}{c}
    \label{eq:edge-set-Qn}
    \{v,v'\} \in \mat{E} \iff \|v-v'\|_2 = 1.
  \end{IEEEeqnarray}
  Thus, by \eqref{eq:only-if},
  \begin{IEEEeqnarray}{rCl}
    \{v,v'\} \in \mat{E}
    & \iff & \|\varphi(v)-\varphi(v')\|_2 = 1
    \\
    & \iff & \bigl\{\varphi(v),\varphi(v')\bigr\} \in \mat{E}
    \quad\forall v, v' \in \mat{V}, 
    \IEEEeqnarraynumspace
  \end{IEEEeqnarray}
  which means that $\varphi \in \const{Aut}(\mat{\Gamma}_{\cube^n})$.
\end{IEEEproof}

\subsection{Proof of Theorem~\ref{thm:Sym-Aut-cube}}

Using Corollary~\ref{coro:proof-thm-Sym-Aut-cube}, we can write
\begin{IEEEeqnarray}{rCl}
  \IEEEeqnarraymulticol{3}{l}{%
    \const{Aut}(\mat{\Gamma}_{\cube^n})
  }\nonumber\\*\;%
  & \cong & \bigl\{\const{g}\in \const{S}_{2^n} \colon
  \|v-v'\|_2 = \|\const{g}v-\const{g}v'\|_2 , \,
  \forall v, v' \in \set{V}(\cube^n) \bigr\}
  \nonumber\\*
  \\
  & = & \const{G}_{\set{V}(\cube^n)}.
  \label{eq:Aut-Qn-iso-group}
\end{IEEEeqnarray}
Furthermore, applying Lemma~\ref{lem:SymP-vertex}, we have
\begin{IEEEeqnarray}{c}
  \const{Sym}_{E_n}(\cube^n) \cong \const{G}_{\set{V}(\cube^n)},
\end{IEEEeqnarray}
and combining this with \eqref{eq:Aut-Qn-iso-group} we finally get
\begin{IEEEeqnarray}{c}
  \const{Sym}_{E_n}(\cube^n) \cong
  \const{Aut}(\mat{\Gamma}_{\cube^n}),
\end{IEEEeqnarray}
concluding the proof.




\end{document}